\documentclass[a4paper,11pt]{article}
\usepackage{jheppub} 
\usepackage[normalem]{ulem}  
\usepackage{cancel}
\allowdisplaybreaks

\definecolor{caribbeangreen}{rgb}{0.0, 0.8, 0.6}
\definecolor{brightmaroon}{rgb}{0.76, 0.13, 0.28}

\newcommand{\De}{{\Delta}}

\newcommand{\al}{{\alpha}}
\newcommand{\bt}{{\beta}}

\newcommand{\de}{{\delta}}
\newcommand{\ep}{{\epsilon}}

\newcommand{\zt}{{\zeta}}
\newcommand{\te}{{\theta}}

\newcommand{\lm}{{\lambda}}

\newcommand{\sig}{{\sigma}}
\newcommand{\vphi}{{\varphi}}

\newcommand{\bsa}{{\boldsymbol{a}}}
\newcommand{\bsb}{{\boldsymbol{b}}}
\newcommand{\bsk}{{\boldsymbol{k}}}
\newcommand{\bsp}{{\boldsymbol{p}}}

\newcommand{\bsep}{{\boldsymbol{\epsilon}}}
\newcommand{\bspd}{{\boldsymbol{\partial}}}

\newcommand{\bsK}{{\boldsymbol{K}}}
\newcommand{\bsX}{{\boldsymbol{X}}}

\newcommand{\bspi}{{\boldsymbol{\pi}}}
\newcommand{\bsPi}{{\boldsymbol{\Pi}}}

\newcommand{\bbD}{{\mathbb{D}}}

\newcommand{\cC}{{\mathcal{C}}}

\newcommand{\cK}{{\mathcal{K}}}

\newcommand{\hL}{\hat{L}}
\newcommand{\hS}{\hat{S}}
\newcommand{\hs}{\hat{s}}

\newcommand{\lan}{{\langle}}

\newcommand{\nn}{{\nonumber}}

\newcommand{\pd}{{\partial}}

\newcommand{\ran}{{\rangle}}

\newcommand{\wh}{\widehat}

\def\bbra{{\langle\kern-2.5pt\langle}}
\def\kket{{\rangle\kern-2.5pt\rangle}}

\title{Momentum space conformal three-point functions of conserved currents and a general spinning operator}

\author[a]{Hiroshi Isono,}
\author[b]{Toshifumi Noumi}
\author[b]{and Toshiaki Takeuchi}

\affiliation[a]{Department of Physics, Faculty of Science, Chulalongkorn University, Bangkok 10330, Thailand}
\affiliation[b]{Department of Physics, Kobe University, Kobe 657-8501, Japan}

\emailAdd{toshi.takeuchi@stu.kobe-u.ac.jp}
\emailAdd{hiroshi.isono81@gmail.com}
\emailAdd{tnoumi@phys.sci.kobe-u.ac.jp}

\abstract{
We construct conformal three-point functions in momentum space with a general tensor and conserved currents of spin $1$ and $2$. While conformal correlators in momentum space have been studied especially in the connection with cosmology, correlators involving a tensor of general spin and scaling dimension have not been studied very much yet. Such a direction is unavoidable when we go beyond three-point functions because general tensors always appear as an intermediate state. In this paper, as a first step, we solve the Ward-Takahashi identities for correlators of a general tensor and conserved currents. In particular we provide their expression in terms of the so-called triple-$K$ integrals and a differential operator which relates triple-$K$ integrals with different indices. For several correlators, closed forms without the differential operator are also found.
}

\preprint{KOBE-COSMO-19-05}

\begin{document}
\setcounter{tocdepth}{2}
\maketitle
\flushbottom

\section{Introduction}
\setcounter{equation}{0}

Recent development in higher dimensional conformal field theory (CFT) (see, e,g.,~\cite{Qualls:2015qjb,Rychkov:2016iqz,Simmons-Duffin:2016gjk,Penedones:2016voo} for review) has boosted research activities in rather new applications to cosmology~\cite{Maldacena:2002vr,Antoniadis:2011ib,Maldacena:2011nz,Creminelli:2012ed,Schalm:2012pi,Mata:2012bx,McFadden:2013ria,Ghosh:2014kba,Bzowski:2012ih,Kundu:2014gxa,Arkani-Hamed:2015bza,Kundu:2015xta,Shukla:2016bnu,Isono:2016yyj,Arkani-Hamed:2018kmz,Goon:2018fyu} and condensed matter physics~\cite{Chowdhury:2012km,Huh:2013vga,Jacobs:2015fiv,Myers:2016wsu,Lucas:2016fju,Lucas:2017dqa}. For example, inflationary correlation functions are highly constrained by conformal symmetry (realized at the future boundary of de Sitter space) and their analytic properties in momentum space are crucial to identify the particle spectrum at the inflationary scale~\cite{Chen:2009zp, Baumann:2011nk, Noumi:2012vr,Arkani-Hamed:2015bza}.
These applications have then pushed forward studies on CFT in momentum space~\cite{Ferrara:1974nf,Bzowski:2012ih,Bzowski:2013sza,Bzowski:2015pba,Bzowski:2015yxv,Coriano:2012hd,Coriano:2013jba,Bzowski:2017poo,Kundu:2014gxa,Arkani-Hamed:2015bza,Kundu:2015xta,Shukla:2016bnu,Coriano:2018bbe,Bzowski:2018fql,Gillioz:2018mto,Coriano:2018bsy,Farrow:2018yni,Skvortsov:2018uru}. However, its understanding is still limited compared to the well-developed position space one, hence further studies on conformal correlators in momentum space are encouraged.

\medskip
The purpose of this paper is to construct three-point functions of conserved currents and a general tensor in momentum space. Recent studies on conformal correlators in momentum space have been triggered by a seminal work~\cite{Bzowski:2013sza} by Bzowski et al, which systematically studied the conformal Ward-Takahasi (WT) identities in momentum space and constructed three-point functions of conserved currents and scalars. For example, three-point functions of primary scalars are given up to an overall normalization factor by~\cite{Ferrara:1974nf,Bzowski:2012ih,Bzowski:2013sza,Bzowski:2015pba,Bzowski:2015yxv}
\begin{align}
\langle \varphi_1(\bsp_1)\varphi_2(\bsp_2)\varphi_3(\bsp_3)\rangle'=J_{0\{0,0,0\}}(p_1,p_2,p_3)\,,
\end{align}
where we introduced $\langle\,\ldots\,\rangle=(2\pi)^d\delta^d(\sum \bsk_i)\langle\,\ldots\,\rangle'$ and the so-called triple-$K$ integral (a more precise definition will be given later):
\begin{align}
J_{N\{k_{1},k_{2},k_{3}\}}(p_{1},p_{2},p_{3})
 & =\int_{0}^{\infty}\frac{dz}{z}\:z^{2d-\Delta_{t}-k_{t}+N}
 \prod_{i=1}^3\left(p_{i}z\right)^{\nu_{i}+k_{i}}K_{\nu_{i}+k_{i}}(p_{i}z)
 \,.
\end{align}
Similarly, three-point functions involving conserved currents of spin $1$ and $2$ and scalars can be written in terms of the triple-$K$ integrals $J_{N\{k_{1},k_{2},k_{3}\}}$ with general indices~\cite{Bzowski:2013sza}.

\medskip

So far, momentum space correlators involving a tensor of general spin and dimension have not been studied very much compared to  position space ones \cite{Costa:2011mg,Sotkov:1976xe,Sotkov:1980qh}. However, they are necessary especially when we go beyond three-point functions because general tensors appear as an intermediate state whatever external operators we consider.
In~\cite{Isono:2018rrb} two of the present authors and a collaborator constructed three-point functions with two scalars and a general tensor. Using the obtained three-point functions, they further constructed a crossing symmetric basis (the Polyakov block) of scalar four-point functions with a general intermediate state, which may be used, e.g., for the Polyakov type bootstrap approach~\cite{Polyakov:1974gs,Sen:2015doa,Gopakumar:2016wkt, Gopakumar:2016cpb,Gopakumar:2018xqi} complementary to the ordinary bootstrap approach\footnote{
Indeed, momentum space was employed in the pioneering work~\cite{Polyakov:1974gs} to make analyticity manifest and utilize dispersion relations to accomplish the bootstrap program in the $\text{O}(N)$ model. It was recently revisited and extended in~\cite{Sen:2015doa}. We hope that our momentum space approach to conformal correlators is helpful for proceeding in this direction.}.
In this paper, along the same line,
 we construct three-point functions involving conserved currents and a general tensor, as a first step toward studies of four-point functions of conserved currents. We provide their expression in terms of triple-$K$ integrals and a differential operator which relates triple-$K$ integrals with different indices. Furthermore, for correlators with no or one conserved current, we can find closed forms without the differential operator. This is based on the fact that correlators can be expanded in triple-$K$ integrals and the differential operators just shift the indices of them.

\medskip
The rest of the paper is organized as follows. In the next section we summarize the WT identities in momentum space to clarify our notation. In Sec.~\ref{sec:OOs}, to elaborate on our methodology to solve the WT identities, we review and expand the story for three-point functions of two scalars and a general tensor~\cite{Isono:2018rrb}. We then generalize it to correlators with a single conserved current as well as a scalar and a general tensor (Sec.~\ref{sec:JJs}) and then to those with two conserved currents and a general tensor (Sec.~\ref{sec:TTs}). Some technical details are collected in appendices.

\section{Ward-Takahashi identities}
\label{sec:Ward-Takahashi-Identities}
\setcounter{equation}{0}

In this section we clarify our notation by briefly summarizing the conformal Ward-Takahashi (WT) identities and conservation laws of currents in momentum space.

\paragraph{Notation}

In this paper we discuss three-point functions with a general spinning operator. A standard technique to handle symmetric traceless tensors $\mathcal{O}_{\mu_{1}\mu_{2}\ldots\mu_{s}}$ is to contract all the tensor indices with a null vector $\boldsymbol{\epsilon}$ called the polarization vector~\cite{Bargmann:1977gy,Costa:2011mg}. We also use the shorthand notation,
\begin{equation}
{\epsilon^{s}.\mathcal{O}}=\epsilon^{\mu_{1}}\epsilon^{\mu_{2}}\cdots\epsilon^{\mu_{s}}\mathcal{O}_{\mu_{1}\mu_{2}\ldots\mu_{s}}\,.\label{eq:polarization_vector}
\end{equation}
For example, we write three-point functions of three symmetric traceless tensors as
\begin{align}
&\langle{\epsilon_{1}^{s_{1}}.O_1}(\boldsymbol{p_{1}}){\epsilon_{2}^{s_{2}}.O_2}(\boldsymbol{p_{2}}){\epsilon_{3}^{s_{3}}.O_3}(\boldsymbol{p_{3}})\rangle
\nn\\
&\quad
=(2\pi)^d\delta^d(\bsp_1+\bsp_2+\bsp_3)
\langle{\epsilon_{1}^{s_{1}}.O_1}(\boldsymbol{p_{1}}){\epsilon_{2}^{s_{2}}.O_2}(\boldsymbol{p_{2}}){\epsilon_{3}^{s_{3}}.O_3}(\boldsymbol{p_{3}})\rangle^{\prime}\,,
\end{align}
where the prime on the r.h.s. implies that the delta function factor for momentum conservation is dropped. We focus on Euclidean correlators in parity invariant $d$-dimensional CFT\footnote{
In $d=3$ there appears degeneracy in tensor structures, which requires a separate argument (see, e.g.,~\cite{Bzowski:2013sza}). We leave the analysis in $d=3$ for future work, focusing on $d\geq4$. }. In the following we mostly use the primed correlator. We also use $s_i$ and $\Delta_i$ for the spin and dimension of $O_i$, respectively.

\paragraph{Conformal WT identities}

We then summarize the conformal WT identities in momentum space. First, the dilatation WT identity is given by
\begin{align}
\left(\sum_{i=1}^3\bsp_i\cdot\bspd_i+2d-\Delta_t\right)
\langle{\epsilon_{1}^{s_{1}}.O_1}(\boldsymbol{p_{1}}){\epsilon_{2}^{s_{2}}.O_2}(\boldsymbol{p_{2}}){\epsilon_{3}^{s_{3}}.O_3}(\boldsymbol{p_{3}})\rangle^{\prime}=0 \,,\label{eq:dilatation_momentum}
\end{align}
where we introduced $\De_t=\De_1+\De_2+\De_3$ and $\bspd_i=\pd/\pd\bsp_i$. On the other hand, the special conformal WT identity parametrized by a transformation parameter $\boldsymbol{b}$ reads (see, e.g.,~\cite{Bzowski:2013sza} for details)
\begin{align}
\boldsymbol{b}\cdot\boldsymbol{K}\,
\langle{\epsilon_{1}^{s_{1}}.O_1}(\boldsymbol{p_{1}}){\epsilon_{2}^{s_{2}}.O_2}(\boldsymbol{p_{2}}){\epsilon_{3}^{s_{3}}.O_3}(\boldsymbol{p_{3}})\rangle^{\prime}=0\,,
\label{eq:SCWT}
\end{align}
where we introduced the differential operator $\bsK$ by
\begin{align}
\boldsymbol{K}&=\boldsymbol{K}_s+\boldsymbol{K}_{\epsilon_1}+\boldsymbol{K}_{\epsilon_2}+\boldsymbol{K}_{\epsilon_3}\,,
\\*
\boldsymbol{b}\cdot\boldsymbol{K}_s&=\sum_{i=1}^{3}\big[\bsb\cdot\bspd_{i}\left(-2\left(\Delta_{i}-d+1\right)+2\bsp_{i}\cdot\bspd_{i}\right)-\left(\bsb\cdot \bsp_{i}\right)\bspd_{i}^{2}\big]\,,
\\*
\boldsymbol{b}\cdot\boldsymbol{K}_{\epsilon_i}&=2(\bsep_{i}\cdot\bspd_{i}){(\bsb\cdot\bspd_{\epsilon_{i}})}-2(\bsb\cdot\bsep_{i})({\bspd_{i}\cdot\bspd_{\epsilon_{i}})}\,.
\end{align}
Here the differential operator $\bspd_{\ep_i}=\pd/\pd\bsep_i$ acts on $\bsep_i$ as if it were unconstrained.
Notice that $\boldsymbol{K}_{\epsilon_i}$ trivially acts on the correlator when $O_i$ is scalar. 

\paragraph{Transverse and trace WT identities}

On top of general symmetric traceless tensors, we consider a spin $1$ conserved current $J_\mu$ and the energy-momentum tensor $T_{\mu\nu}$. For simplicity, we focus on Abelian symmetries, even though extension to non-Abelian symmetries is straightforward. In classical theory, they satisfy the transverse WT identities,
\begin{align}
\label{transverse_WT}
p_\mu J^\mu(\bsp)=0\,,
\quad
p_\mu T^{\mu\nu}(\bsp)=0\,,
\end{align}
and the trace WT identity,
\begin{align}
\label{trace_WT}
T_\mu{}^\mu(\bsp)=0\,,
\end{align}
up to equations of motion. In quantum theory, these relations are corrected by local terms in general. However, there are no such corrections for the correlators we consider in this paper\footnote{
For example, let us consider three-point functions of one scalar, one conserved current and a general tensor. The WT identity is then stated as
 \begin{align}
 {\bsp_2}_\mu \langle \varphi(\bsp_1) J^\mu(\bsp_2) \epsilon_3^s.O(\bsp_3)\rangle^\prime
 =-q_1\langle \varphi(\bsp_1)\epsilon_3^s.O(-\bsp_1)\rangle^\prime 
 -q_3\langle \varphi(\bsp_3)\epsilon_3^s.O(-\bsp_3)\rangle^\prime\,,
 \end{align}
 where $q_1\text{ and }q_3$ are the $\text{U}(1)$ charges of the operators $\varphi \text{ and } \mathcal{O}_{\mu_{1}\mu_{2}\ldots\mu_{s}}$. The local terms in the right-hand side vanish since they are proportional to two-point functions of two different operators. The same argument applies to the other correlators studied in this paper. Note that we follow the definition of three-point functions  in~\cite{Osborn:1993cr}. On the other hand, if we employ another definition used in \cite{Bzowski:2013sza}, there may appear local terms. See, e.g.,~Sec. 2.3 of \cite{Bzowski:2018fql} for details of the two different conventions.}.
 Also, the trace WT identity may be modified when there exists a trace anomaly (see~\cite{Bzowski:2015pba,Bzowski:2018fql,Bzowski:2017poo} for details). For technical simplicity, we assume that spinning operators other than the conserved currents in three-point functions have generic scaling dimensions, so that there appears no anomalous term in our analysis. Under these assumptions, the transverse and trace WT identities are simply equivalent to the classical ones~\eqref{transverse_WT}-\eqref{trace_WT}.

\section{Two scalars and a general tensor}
\label{sec:OOs}
In the following sections we solve the conformal WT identities for correlators involving conserved currents and a general tensor. To elaborate on our methodology, we first review and expand the story for three-point functions of two scalars and a general tensor~\cite{Isono:2018rrb}. We then generalize it to correlators involving one conserved current in the next section. Correlators involving two identical conserved currents will be given in Sec.~\ref{sec:TTs}.

\paragraph{General ansatz}
Let us consider three-point functions,
\begin{align}
\lan \vphi_1(\bsp_1)\vphi_2(\bsp_2)\ep_3^s.O(\bsp_3) \ran'\,,
\end{align}
of two scalars $\vphi_1,\vphi_2$ and a spin $s$ symmetric traceless tensor $O$. Throughout the paper, we assume that the tensor $O$ has a nonzero spin $s\neq0$ and a generic dimension $\Delta_3$, and thus is not conserved. From now on, we write three-point functions as functions of $\bsp_1$ and $\bsp_2$ without explicit dependence on $\bsp_3$. In other words we use $\bsp_3=-(\bsp_1+\bsp_2)$ to remove the $\bsp_3$-dependence. The special conformal WT identity then reads
\begin{align}
\label{ssO-scWT}
\bsb\cdot\bsK_s \lan \vphi_1(\bsp_1)\vphi_2(\bsp_2)\ep_3^s.O(\bsp_3) \ran' = 0\,,
\end{align}
where notice that $\bsK_\ep$ does not appear since $\vphi_1,\vphi_2$ are scalar and there is no explicit $\bsp_3$-dependence. To solve the WT identities, we employ the following general ansatz\footnote{
The $1\leftrightarrow2$ exchange symmetry is obscured by the ansatz~\eqref{ssO-ansatz} because the correlator is expanded in $\xi=\bsep_3\cdot\bsp_2$ and $\zeta=\bsep_3\cdot(\bsp_1+\bsp_2)$. To make it manifest, it is more convenient to employ the expansion in $\bsep_3\cdot(\bsp_1\pm\bsp_2)$. See~\cite{Isono:2018rrb} for details. However, it turns out that the ansatz~\eqref{ssO-ansatz} is more useful to derive a closed form such as Eq.~\eqref{closed_OOS}.
}:
\begin{align}
\label{ssO-ansatz}
\lan \vphi_1(\bsp_1)\vphi_2(\bsp_2)\ep_3^s.O(\bsp_3) \ran'
= \sum_{n=0}^s \frac{1}{n!} \xi^n \zt^{s-n} A_n(p_1,p_2,p_3) \,,
\end{align}
where $p_i=|\bsp_i|$ and we introduced
\begin{align}
\label{xizeta}
\xi=\bsep_3\cdot\bsp_2\,, \quad \zt=\bsep_3\cdot(\bsp_1+\bsp_2)\,.
\end{align}
Note also that $p_3$ should be understood as $p_3=|\bsp_1+\bsp_2|$.
The dilatation WT identity then implies that the function $A_n(p_1,p_2,p_3)$ must be homogeneous of degree $\De_t-2d-s$,
\begin{align}
\label{ssO-An-homog}
A_n(\lm p_1,\lm p_2,\lm p_3) = \lm^{\De_t-2d-s} A_n(p_1,p_2,p_3)\,,
\end{align}
where $\De_t=\De_1+\De_2+\De_3$. 

\subsection{Reformulating WT identities}

Our goal is now to determine the function $A_n$ by solving the WT identity \eqref{ssO-scWT} with the ansatz \eqref{ssO-ansatz} and the condition \eqref{ssO-An-homog}. Let us first investigate the $\bsb$-dependence of the left hand side of the WT identity \eqref{ssO-scWT}.
Since it is Lorentz scalar, it is generally of the form,
\begin{align}
\label{SCWT_schematic}
\bsb\cdot\bsK_s \lan \vphi_1(\bsp_1)\vphi_2(\bsp_2)\ep_3^s.O(\bsp_3) \ran' 
= (\bsb\cdot\bsp_1) P_1 + (\bsb\cdot\bsp_2) P_2 + (\bsb\cdot\bsep_3) R\,.
\end{align}
A concrete form of $P_i$ and $R$ is given shortly.
The special conformal WT identity \eqref{ssO-scWT} is then equivalent to $P_1=P_2=R=0$. 
Below we rewrite them in terms of the differential operators with respect to $\xi,\zt,p_1,p_2,p_3$ utilizing various formulae
about $\bsK_s$ summarized in Appendix~\ref{appsec:Ks}.

\medskip
Let us start with the equations, $P_1=0$ and $P_2=0$. Using the formulae \eqref{albt-P1}-\eqref{albt-P2} with $\al=\bt=0$, we may reduce them into the form,
\begin{align}
0 &= [\cK_1(\nu_1) - \cK_3(\nu_3) - 2p_3^{-2}\te_{3}\zt\pd_\zt] ~ \sum_{n=0}^s \frac{1}{n!} \xi^n \zt^{s-n} A_n(p_1,p_2,p_3) \,, 
\label{ssO-P1} \\
0 &= [\cK_2(\nu_2) - \cK_3(\nu_3) - 2p_3^{-2}\te_{3}\zt(\pd_\xi+\pd_\zt)] ~  
\sum_{n=0}^s \frac{1}{n!} \xi^n \zt^{s-n} A_n(p_1,p_2,p_3) \,, \label{ssO-P2}
\end{align}
where we introduced the Euler operator $\te_i=p_i\pd_{p_i}$ for the momentum $p_i$.
The differential operators $\cK_i(\nu_i)$ with respect to $p_i$ are defined as
\begin{align}
\cK_i(\nu_i) = p_i^{-2}\te_{i}(\te_{i}-2\nu_i) \,,
\end{align}
where $\nu_i=\De_i-(d/2)$.
A remark is that the variables $p_1,p_2,p_3,\xi,\zt$ of the differential operators have to be regarded as independent variables.

\medskip
On the other hand, we may express the equation $R=0$ by using the formula \eqref{albt-R} with $\al=\bt=0$ as
\begin{align}
\label{ssO-R}
0 = [\te_x(\te_x+\Xi-1) - x(\te_x-s)(\te_x+\De_3-1)] \sum_{n=0}^s \frac{1}{n!}x^nA_n(p_1,p_2,p_3)\,,
\end{align}
where $x=\xi/\zt$ and $\te_x=x\pd_x$. The differential operator $\Xi$ is defined by
\begin{align}
\label{Xi_def}
\Xi = \frac{1}{2}\bigg( \De_1-\De_2+\De_3-s-\te_{1}+\te_{2}-\frac{p_1^2-p_2^2}{p_3^2}\te_{3} \bigg)\,.
\end{align}
Notice that $\Xi$ commutes with $x$ and $\theta_x$ in particular.

\subsection{Solving WT identities}

We proceed to solving the WT identities. Below we first use the WT identities \eqref{ssO-P1}-\eqref{ssO-P2} to determine $A_s$. Using this $A_s$ as an initial condition for a recursion relation derived from Eq.~\eqref{ssO-R}, we provide an expression for the other $A_n$. 

\subsubsection*{Initial condition $A_s$}
In general, the WT identities \eqref{ssO-P1}-\eqref{ssO-P2} are rather complicated relations among $A_n$ with different $n$. However, their $\mathcal{O}(\zeta^0)$ terms provide differential equations containing $A_s$ only:
\begin{align}
\label{ssO-As-eqs}
0 &= [\cK_1(\nu_1) - \cK_3(\nu_3)]A_s(p_1,p_2,p_3) 
= [\cK_2(\nu_2) - \cK_3(\nu_3)]A_s(p_1,p_2,p_3) \,.
\end{align}
Together with the dilatation WT identity~\eqref{ssO-An-homog}, we can solve these equations in terms of the so-called triple-$K$ integrals as
\begin{align}
\label{ssO-As}
A_s = \cC_A J_{s\{0,0,0\}}(p_1,p_2,p_3) \,,
\end{align}
where $\cC_A$ is an undetermined overall coefficient. We also introduced\footnote{
Note that the integral~\eqref{eq:triple-K_integral} is convergent only when $|{\rm Re}\,\nu_1|+|{\rm Re}\,\nu_2|+|{\rm Re}\,\nu_3|<s+\frac{d}{2}$. Otherwise, there appears a singularity near $z=0$ and we need to perform analytic continuation~\cite{Bzowski:2015pba,Bzowski:2015yxv}, which may be carried out, e.g., by introducing the Pochhammer contour.}
\begin{align}
J_{N\{k_{1},k_{2},k_{3}\}}(p_{1},p_{2},p_{3})
 & =\int_{0}^{\infty}\frac{dz}{z}\:z^{2d-\Delta_{t}-k_{t}+N}
 \prod_{i=1}^3\left(p_{i}z\right)^{\nu_{i}+k_{i}}K_{\nu_{i}+k_{i}}(p_{i}z)
 \,,
 \label{eq:triple-K_integral}
\end{align}
where $K_\nu(z)$ is the Bessel function of the second kind and $k_t=k_1+k_2+k_3$. See~Appendix~\ref{app:tripleK} for derivation of~\eqref{ssO-As}. There we also summarize various properties of the triple-$K$ integral~\eqref{eq:triple-K_integral}.

\subsubsection*{Recursion relations for $A_n$}

We would then like to determine the other $A_n$. For this purpose, it is convenient to use the other WT identity \eqref{ssO-R}, which provides a recursion relation,
\begin{align}
(\Xi+n)A_{n+1} = (-s+n)(\De_3-1+n)A_n\,. \label{00s-recursion}
\end{align}
Here we emphasize that the differential operator $\Xi$ is acting only on $A_{n+1}$, hence we can express $A_n$ with a lower $n$ as a derivative of higher $A_n$. Indeed, it is easy to find\footnote{Eq.~\eqref{ssO-R} is nothing but the hypergeometric differential equation, hence its solution is given by
\begin{align}
\sum_{n=0}^s \frac{1}{n!}x^nA_n(p_1,p_2,p_3) \propto {}_2F_1\left( -s,\De_3-1; \Xi; x \right)\,,
\end{align}
where we chose a polynomial solution in $x$. The proportionality constant is fixed by $A_s=\cC_AJ_{s(0,0,0)}$.}
\begin{align}
\label{An_OOs}
A_n = \frac{(\Xi+n)_{s-n}}{(-s+n)_{s-n}(\De_3-1+n)_{s-n}}A_s
=\cC_A\frac{(\Xi+n)_{s-n}}{(-s+n)_{s-n}(\De_3-1+n)_{s-n}}J_{s\{0,0,0\}}
\, ,
\end{align}
where $\displaystyle(x)_n=x(x+1)\cdots (x+n-1)$ is the shifted factorial (also dubbed the Pochhammer symbol). Even though this already provides a compact expression, it is useful to explore an expression without differential operators. As given in Eq.~\eqref{Xi_on_tripleK}, the differential operator $\Xi$ relates triple-$K$ integrals with different indices. Together with~\eqref{k3_down}, we can always expand $A_n$ by triple-$K$ integrals with a fixed $k_3$ index as
\begin{align}
\label{tripleK_expansion}
A_n=\sum_{k_1,k_2\geq0}a_{n\{k_1,k_2\}}J_{n+k_1+k_2\{k_1,k_2,n-s\}}\,.
\end{align}
The coefficients $a_{n\{k_1,k_2\}}$ can then be determined algebraically by using Eq.~\eqref{Xi_on_tripleK}. 
The result is that the matrix components with $k_1>0$ are all zero and the non-zero components are given by\footnote{
In Appendix~\ref{app:another} we provide an alternative derivation of the coefficients~\eqref{closed_OOS}, which is useful when working on correlators with at most two spinning operators. However, it turns out to be not straightforward to apply it to correlators of three spinning operators. On the other hand, the algebraic calculation presented in this section provides a general framework applicable to any correlator.}

\begin{align}
a_{s-n\{0,k\}}=\cC_A\frac{2^{n-k}(1\!-\!\tfrac{s-\De_1+\De_2+\De_3}{2})_{n-k}(1+\tfrac{d-s-\De_t}{2})_{n-k}}{k!(n-k)!(2-\Delta_3-s)_{n-k}}
\quad
(0\leq k\leq n)\,.
\label{closed_OOS}
\end{align}

\subsubsection*{Residual WT identities}

So far, we have not checked yet if the full three-point function with~\eqref{An_OOs} satisfies the WT identities \eqref{ssO-P1}-\eqref{ssO-P2} (only a part of which was used to determine $A_s$). From the position space results, we know that there is only one free parameter $\cC_A$, hence Eq.~\eqref{An_OOs} should be consistent with all the WT identities. To conclude this section, we explicitly show that it is indeed the case.

\medskip
First, the WT identities~\eqref{ssO-P1}-\eqref{ssO-P2} can be expressed in terms of $A_n$ as
\begin{align}
\label{1-3_expand}
\left[\cK_1(\nu_1) - \cK_3(\nu_3)\right]A_n&=2(s-n)p_3^{-2}\theta_3A_n\,,
\\*
\label{2-3_expand}
\left[\cK_2(\nu_2) - \cK_3(\nu_3)\right]A_n&=2(s-n)p_3^{-2}\theta_3A_n+2p_3^{-2}\theta_3A_{n+1}\,.
\end{align}
As we mentioned earlier, $A_n$ can be expanded by triple-$K$ integrals as Eq.~\eqref{tripleK_expansion}. From the formulae~\eqref{Euler_on_tripleK}-\eqref{K_on_tripleK}, we find that the differential operators in Eqs.~\eqref{1-3_expand}-\eqref{2-3_expand} act on each triple-$K$ integral as
\begin{align}
\left[\cK_1(\nu_1) - \cK_3(\nu_3)\right]J_{N\{k_1,k_2,k_3\}}&=-2k_1J_{N+1\{k_1-1,k_2,k_3\}}+2k_3J_{N+1\{k_1,k_2,k_3-1\}
}\,,
\\
p_3^{-2}\theta_3J_{N\{k_1,k_2,k_3\}}&=-J_{N+1\{k_1,k_2,k_3-1\}}\,.
\end{align}
Then, we may translate the WT identities~\eqref{1-3_expand}-\eqref{2-3_expand} into algebraic relations among the coefficients $a_{n\{k_1,k_2\}}$ as
\begin{align}
k_1a_{n\{k_1,k_2\}}=0\,,
\quad
k_2a_{n\{k_1,k_2\}}=a_{n+1\{k_1,k_2-1\}}\,,
\end{align}
which are indeed satisfied by our solution Eq.~\eqref{closed_OOS}. The closed form~\eqref{tripleK_expansion}-\eqref{closed_OOS} without differential operators and the explicit check of the full WT identities are new results of the present paper.

\paragraph{Summary of the section}
In this way, the WT identities can be reformulated into several sets of differential equations. In the present case, some of them are used to derive recursion relations for functional coefficients in the decomposition~\eqref{ssO-ansatz}, while the others provided their initial conditions. This step specifies three-point functions up to a free parameter in terms of triple-$K$ integrals and a differential operator relating triple-$K$ integrals with different indices. We can also expand it by triple-$K$ integrals without using differential operators, whose coefficients can be calculated algebraically. This expression translates the WT identities into algebraic relations among these coefficients. While we succeeded in providing a closed form~\eqref{closed_OOS} of the coefficients in this section, it is not easy for more complicated correlators such as the ones discussed in Sec.~\ref{sec:TTs} to derive a closed form for general spin $s$ of the tensor. However, our algebraic approach is still tractable enough for concrete problems for a given spin $s$, e.g., with the help of computer software.

\section{One scalar, one conserved current and a general tensor}
\label{sec:JJs}

We extend the argument in the previous section to solve the conformal WT identities for correlators with a scalar, a conserved current, and a general tensor.

\subsection{Spin 1 conserved current $J_\mu$}

Let us begin by three-point functions of a scalar $\vphi$, a spin $1$ conserved current $J_\mu$, and a spin $s$ symmetric traceless operator $O$:
\begin{align}
\lan \vphi(\bsp_1) \ep_2.J(\bsp_2) \ep_3^s.O(\bsp_3) \ran' \,,
\end{align}
where the helicity vector $\bsep_3$ is null to respect the tracelessness of $O$. Before solving the conformal WT identities, it is convenient to impose the conservation law $\partial_\mu J^\mu=0$ first, which is achieved by parameterizing three-point functions as
\begin{align}
& \lan \vphi(\bsp_1) \ep_2.J(\bsp_2) \ep_3^s.O(\bsp_3) \ran' = 
(\bsep_2\cdot\bspi_2\cdot\bsp_1) A + (\bsep_2\cdot\boldsymbol{\pi}_2\cdot\bsep_3) B \,.
\end{align}
Here we introduced the transverse projector,
\begin{align}
(\pi_i)_{\mu\nu} = \de_{\mu\nu}-\frac{(p_i)_\mu(p_i)_\nu}{p_i^2} \label{pi}\,,
\end{align}
and $\bsep_2\cdot\boldsymbol{\pi}_2\cdot\bsp_1=(\ep_2)^\mu(\pi_2)_{\mu\nu}(p_1)^\nu$ for example. We further expand $A$ and $B$ as
\begin{align}
\label{OJS_ansatz}
A =  \sum_{n=0}^s \frac{1}{n!} \xi^n\zt^{s-n} A_n(p_1,p_2,p_3) \,, \quad
B = \sum_{n=0}^{s-1} \frac{1}{n!} \xi^n\zt^{s-1-n} B_n(p_1,p_2,p_3)\,,
\end{align}
where $A_n$ and $B_n$ are scalar functions of $p_i$, and $\xi=\bsep_3\cdot\bsp_2$ and $\zt=\bsep_3\cdot(\bsp_1+\bsp_2)$ as before.

\subsubsection{Reformulating WT identities}

Just as we did in the previous section, we reformulate the conformal WT identities into differential equations for $A_n$ and $B_n$. First, the dilatation WT identity simply gives the homogeneity conditions on $A_n$ and $B_n$ as
\begin{align}
A_n(\lm p_1,\lm p_2,\lm p_3) &= \lm^{\De_t-2d-s-1} A_n(p_1,p_2,p_3) \,,\label{01s-homoA}\\*
B_n(\lm p_1,\lm p_2,\lm p_3) &= \lm^{\De_t-2d-s+1} B_n(p_1,p_2,p_3) \,.\label{01s-homoB}
\end{align}
We next consider the special conformal WT identities:
\begin{align}
0 = (\bsb\cdot\bsK_s + \bsb\cdot\bsK_{\ep_2}) 
\lan \vphi(\bsp_1) \epsilon_2.J(\bsp_2) \ep_3^s.O(\bsp_3) \ran' \,.
\end{align}
In the previous section we split the WT identities for the two scalar case into three differential equations based on how the transformation parameter $\bsb$ is contracted. See Eq.~\eqref{SCWT_schematic}. In the present case we find seven equations as below.

\medskip
Let us first summarize how the differential operator $\bsK_s$ acts on $A$ and $B$. Similarly to the calculation in the previous section, we find
\begin{align}
(\bsb\cdot\bsK_s)A &= (\bsb\cdot\bsp_1)P_1^{(-1,0)}A+(\bsb\cdot\bsp_2)P_2^{(-1,0)}A+(\bsb\cdot\bsep_3)R^{(-1,0)}A\,, \label{X-decomp} \\
(\bsb\cdot\bsK_s)B &= (\bsb\cdot\bsp_1)P^{(1,1)}_1B+(\bsb\cdot\bsp_2)P^{(1,1)}_2B+(\bsb\cdot\bsep_3)R^{(1,1)}B\,, \label{Y-decomp}
\end{align}
where $P^{(\alpha,\beta)}_{1}$, $P^{(\alpha,\beta)}_{2}$, and $R^{(\alpha,\beta)}$ are differential operators defined by Eqs.~\eqref{albt-P1}, \eqref{albt-P2}, and \eqref{albt-R}. 
In this language, we obtain four differential equations from the four terms proportional to $\bsb\cdot\bsp_i$ ($i=1,2$) as
\begin{align}
(\bsep_2\cdot\bspi_2\cdot\bsp_1)(\bsb\cdot\bsp_1) &: \quad 
0 = \left( P_1^{(-1,0)} + 2p_3^{-2}\te_{3} \right)A \,, \label{e2p1bp1} \\
(\bsep_2\cdot\bspi_2\cdot\bsp_1)(\bsb\cdot\bsp_2) &: \quad 
0 =  \left( P_2^{(-1,0)} + 2p_3^{-2}\te_{3} \right)A \,. \label{e2p1bp2} \\
(\bsep_2\cdot\bspi_2\cdot\bsep_3)(\bsb\cdot\bsp_1) &: \quad 
0 =  P^{(1,1)}_1 B+ 2\pd_\xi A \,, \label{e2e3bp1} \\
(\bsep_2\cdot\bspi_2\cdot\bsep_3)(\bsb\cdot\bsp_2) &: \quad 
0 =  P^{(1,1)}_2 B\,. \label{e2e3bp2}
\end{align}
There are also two terms proportional to $\bsb\cdot\bsep_3$, which lead to
\begin{align}
(\bsep_2\cdot\bspi_2\cdot\bsep_3)(\bsb\cdot\bsep_3) &: \quad 
0 = \left( R^{(1,1)} + 2\pd_\xi+2\pd_\zt \right)B \,, \label{e2e3be3} \\*
(\bsep_2\cdot\bspi_2\cdot\bsp_1)(\bsb\cdot\bsep_3) &: \quad 
0 = \left( R^{(-1,0)} + 2\pd_\zt \right)A  + 2p_3^{-2}\te_{3}B \,. \label{e2p1be3} 
\end{align}
Finally,  the term proportional to $\bsep_2\cdot\bspi_2\cdot\bsb$ gives
\begin{align}
(\bsep_2\cdot\bspi_2\cdot\bsb) &: \quad 0 =  \left[ 
-(\De_1-d)+(\xi-\zt)\pd_\xi+\te_{1}+\frac{\bsp_1\cdot\bsp_2}{p_2^2}(d-2-\te_{2})
\right]A \nn\\
&\qquad\qquad +\left[\frac{\xi}{p_2^2}(d-2-\te_{2})-\frac{\zt}{p_3^2}\te_{3}\right]B \,. \label{e2b} 
\end{align}
For later convenience, we classify these identities into the following two:
First, we call the identities~\eqref{e2p1bp1}-\eqref{e2p1be3} associated with $\bsb\cdot\bsp_1$, $\bsb\cdot\bsp_2$, and $\bsb\cdot\bsep_3$ the primary WT identities, following the terminology of~\cite{Bzowski:2013sza}. They are used to determine a functional form of correlators up to several free parameters. On the other hand, we call the identity~\eqref{e2b} associated with $\bsep_2\cdot\bspi_2\cdot\bsb$ the secondary WT identity, which provides a constraint on the free parameters.

\subsubsection{Solving WT identities}

We now proceed to solving the seven differential equations \eqref{e2p1bp1}-\eqref{e2b}. Our strategy for this problem is the following: We start with the primary WT identities. As in the two scalar case, we use the first four equations \eqref{e2p1bp1}-\eqref{e2e3bp2} to determine $A_s$ and $B_{s-1}$. We then use Eqs.~\eqref{e2e3be3}-\eqref{e2p1be3} to find recursion relations for $A_n$ and $B_n$, which specify the form of $A_n$ and $B_n$ up to two free parameters. Finally, we use the secondary WT identity to provide a relation between the two. Afterwards, we are left with a single free parameter, which is consistent with the position space result~\cite{Costa:2011mg}.

\subsubsection*{Initial conditions from primary WT identities}

Let us first use the $\mathcal{O}(\zeta^0)$ terms of Eqs.~\eqref{e2p1bp1}-\eqref{e2e3bp2},
\begin{align}
0=& \left[K_1(\nu_1)-K_3(\nu_3)\right]A_s\,,\label{01s-primary1}\\
0=& \left[K_2(\nu_2)-K_3(\nu_3)\right]A_s\,,\label{01s-primary2}\\
0=& \left[K_1(\nu_1)-K_3(\nu_3)\right]B_{s-1}+2A_s\,,
\label{01s-primary3}\\
0=& \left[K_2(\nu_2)-K_3(\nu_3)\right]B_{s-1}\,,\label{01s-primary4}
\end{align}
to determine the initial conditions, $A_s$ and $B_{s-1}$. Under the homogeneity condition~\eqref{01s-homoA}, we solve Eqs.~\eqref{01s-primary1}-\eqref{01s-primary2} in terms of the triple-$K$ integral as
\begin{align}
A_s=\cC_AJ_{s+1\{0,0,0\}}(p_1,p_2,p_3)\,,
\end{align}
where $\cC_A$ is a free parameter. Next, we solve the other two equations. To find a particular solution for Eqs.~\eqref{01s-primary3}-\eqref{01s-primary4}, it is convenient to employ the ansatz,
\begin{align}
B_{s-1}=bJ_{s-1+k_t\{k_1,k_2,k_3\}}
\end{align}
with a constant $b$, where $k_t=k_1+k_2+k_3$. Note that it satisfies the homogeneity condition~\eqref{01s-homoB}. Using Eq.~\eqref{K_on_tripleK}, we may reduce Eqs.~\eqref{01s-primary3}-\eqref{01s-primary4} to the form,
\begin{align}
0&=-2bk_1J_{s+k_t\{k_1-1,k_2,k_3\}}+2bk_3J_{s+k_t\{k_1,k_2,k_3-1\}}+2\cC_AJ_{s+1\{0,0,0\}}\,,
\\*
0&=-2bk_2J_{s+k_t\{k_1,k_2-1,k_3\}}+2bk_3J_{s+k_t\{k_1,k_2,k_3-1\}}\,,
\end{align}
which can be solved, e.g., by $(b,k_1,k_2,k_3)=(\cC_A,1,0,0)$. Adding the homogeneous solution $J_{s-1\{0,0,0\}}$, we find the general solution,
\begin{align}
B_{s-1}=\cC_AJ_{s\{1,0,0\}}+\cC_B J_{s-1\{0,0,0\}}\,,
\end{align}
with a free parameter $\cC_B$.

\subsubsection*{Recursion relations from primary WT identities}

Next, Eqs.~\eqref{e2e3be3}-\eqref{e2p1be3} can be thought of as recursion relations for $A_n$ and $B_n$:
\begin{align}
0=&\left(\Xi+\frac{3}{2}+n\right)B_{n+1}-(-s+1+n)(\Delta_3+n)B_n\,,\label{01s-Bn}\\*
0=&\left(\Xi-\frac{1}{2}+n\right)A_{n+1}-(-s+n)(\Delta_3-1+n)A_n+p_3^{-2}\te_3B_n\,.
\label{01s-An}
\end{align}
The first equation has the same form as \eqref{00s-recursion}, hence its solution is
\begin{align}
\nonumber
B_n&=\frac{(\Xi+n+\frac{3}{2})_{s-1-n}}{(-s+1+n)_{s-1-n}(\De_3+n)_{s-1-n}}B_{s-1}
\\
&=\frac{(\Xi+n+\frac{3}{2})_{s-1-n}}{(-s+1+n)_{s-1-n}(\De_3+n)_{s-1-n}}\left(
\cC_AJ_{s\{1,0,0\}}+\cC_B J_{s-1\{0,0,0\}}
\right)\,.
\end{align}
Similarly, the solution for Eq.~\eqref{01s-An} is given by
\begin{align}
\nonumber
A_n&=\frac{(\Xi+n-\frac{1}{2})_{s-n}}{(-s+n)_{s-n}(\De_3-1+n)_{s-n}}\cC_AJ_{s+1\{0,0,0\}}\\
&\quad
+\sum_{t=0}^{s-1-n}\frac{(\Xi+n-\frac{1}{2})_t}{(-s+n)_{t+1}(\De_3-1+n)_{t+1}}p_3^{-2}\te_3B_{n+t}\,.
\end{align}

\subsubsection*{A closed form without differential operators}
Just as the two scalar case, it may be convenient to find expressions without the differential operators.
Combining Eq.~\eqref{Xi_on_tripleK} with Eq.~\eqref{k3_down}, we can expand $A_n$ and $B_n$ in triple-$K$ integrals with a fixed $k_3$ index as
\begin{align}
A_n&=\sum_{k_1,k_2\geq0}a_{n\{k_1,k_2\}}J_{n+k_1+k_2+1\{k_1,k_2,n-s\}}\,,
\\
B_n&=\sum_{k_1,k_2\geq0}b_{n\{k_1,k_2\}}J_{n+k_1+k_2\{k_1,k_2,n-s+1\}}\,.
\end{align}
Using Eq.~\eqref{Xi_on_tripleK}, we can compute the coefficients $a_{n\{k_1,k_2\}},b_{n\{k_1,k_2\}}$ algebraically. The result is summarized as follows: 
The coefficients $a_{n\{k_1,k_2\}}$ with $k_1>0$ and $b_{n\{k_1,k_2\}}$ with $k_1>1$ all vanish, and the non-zero coefficients are given by 
\begin{align}
a_{s-n\{0,k\}}&= b_{s-1-n\{1,k\}} \,, 
\\*
b_{s-1-n\{0,k\}}&=\cC_B\frac{2^{n-k}(\frac{1}{2}+\tfrac{\De_1-\De_2-\De_3-s}{2})_{n-k}(\frac{3}{2}+\frac{d-s-\De_t}{2})_{n-k}}{k!(n-k)!(2-\Delta_3-s)_{n-k}}\,,
\\*
b_{s-1-n\{1,k\}}&=
\cC_A\frac{2^{n-k}(\frac{3}{2}+\tfrac{\De_1-\De_2-\De_3-s}{2})_{n-k}(\frac{1}{2}+\frac{d-s-\De_t}{2})_{n-k}}{k!(n-k)!(2\!-\!\Delta_3\!-\!s)_{n-k}}
\nonumber
\\*
\label{b1k_OJS}
&\quad
-\cC_B\frac{2^{n-k-1}(\frac{3}{2}+\tfrac{\De_1-\De_2-\De_3-s}{2})_{n-k-1}(\frac{3}{2}+\frac{d-s-\De_t}{2})_{n-k-1}}{k!(n-k-1)!(2\!-\!\Delta_3\!-\!s)_{n-k}}
\,,
\end{align}
where $0 \leq k \leq n$ (the second line of Eq.~\eqref{b1k_OJS} is interpreted as zero for $n=k$). These coefficients satisfy the identities~\eqref{e2p1bp1}-\eqref{e2e3bp2} out of the primary WT identities as they satisfy the relations~\eqref{01s-initial-conseq1}-\eqref{01s-initial-conseq2} that are equivalent to the four identities.

\subsubsection*{Secondary WT identity}
We have determined three-point functions $\lan \vphi(\bsp_1) \ep_2.J(\bsp_2) \ep_3^s.O(\bsp_3) \ran' $ up to the two free parameters $\cC_A$ and $\cC_B$. Finally, we solve Eq.~\eqref{e2b} and  reduce the number of parameters to one. We again focus on its $\mathcal{O}(\zeta^0)$ terms. By taking the zero-momentum limit, $\bsp_3\rightarrow 0$, triple-$K$ integrals reduce to monomials of $p=p_1=p_2$ and thus Eq.~\eqref{e2b} is simplified as
\begin{align}
0&=\cC_A \Bigr[-j_{s+2\{0,0,0\}}-j_{s+2\{0,-1,0\}}+(-\Delta_1+s+2)j_{s+1\{0,0,0\}} \nn\\*
&\quad
+s(d-2)j_{s\{1,0,0\}}+sj_{s+1\{1,-1,0\}}\Bigr]+s\cC_B\Bigr[(d-2)j_{s-1\{0,0,0\}}+j_{s\{0,-1,0\}}\Bigr]\,,
\end{align}
where $j_{N\{k_1,k_2,k_3\}}$ is a numerical number given in Eq.~\eqref{zero-p3}. The general solution is somewhat complicated, so that we provide two illustrative examples.
For example, for a scalar with $\Delta_1=4,d=5$ dual to a $6$D bulk scalar with a conformal mass and a spinning operator with $s=2,\Delta_3=\frac{11}{2}$, we have
\begin{align}
\cC_B=-\frac{3245}{656}\cC_A\,.
\end{align}
Also, for a scalar with $\Delta_1=5, d=5$ dual to a $6$D massless bulk scalar and a spinning operator with $s=2,\Delta_3=\frac{11}{2}$, we find
\begin{align}
\cC_B=-\frac{1073}{656}\cC_A\,.
\end{align}

\bigskip

\subsection{Energy-momentum tensor $T_{\mu\nu}$}

We next consider three-point functions of a scalar $\phi$, the energy-momentum tensor $T^{\mu\nu}$, and a general tensor:
\begin{align}
\lan \vphi(\bsp_1) \ep_2^{2}.T(\bsp_2) \ep_3^s.O(\bsp_3) \ran' \,.\label{TTS}
\end{align}
The conservation law $\partial_\mu T^{\mu\nu}=0$ and the traceless condition $T^\mu_\mu=0$ result in the ansatz,
\begin{align}
\lan \vphi(\bsp_1) \ep_2^{2}.T(\bsp_2) \ep_3^s.O(\bsp_3) \ran' =
(\bsep_2^2\cdot\bsPi_2\cdot\bsp_1^2)  A +(\bsep_2^2\cdot\bsPi_2\cdot\bsp_1\bsep_3) B +(\bsep_2^2\cdot\bsPi_2\cdot\bsep_3^2) C\,,\label{OTS_ansatz} \,
\end{align}
where we introduced the transverse-traceless projector,
\begin{align}
(\Pi_i)_{\mu\nu\rho\sigma}=\frac{1}{2}\left\{(\pi_i)_{\mu\rho} (\pi_i)_{\nu\sigma}+(\pi_i)_{\mu\sigma} (\pi_i)_{\nu\rho}\right\}-\frac{1}{d-1}(\pi_i)
_{\mu\nu}(\pi)_{i\rho\sigma}\,,
\end{align}
with $\pi_i$ being the transverse projector defined in Eq.~\eqref{pi}. Here we used a shorthand notation, e.g., $\bsep_2^2\cdot\boldsymbol{\Pi}_2\cdot\bsp_1\bsep_3 = (\ep_2)^\mu(\ep_2)^\nu(\Pi_2)_{\mu\nu\alpha\beta}(\epsilon_3)^{\alpha}(p_1)^{\beta}$, for the tensorial contraction of the transverse-traceless projector. Note that since the last term in Eq.~\eqref{OTS_ansatz} requires two or more $\bsep_3$, we have $C=0$ for $s=1$. We also parameterize $A$, $B$, and $C$ as
\begin{align}
\label{OTS_ansatz1}
&A =  \sum_{n=0}^s \frac{1}{n!} \xi^n\zt^{s-n} A_n(p_1,p_2,p_3)\,, \\
&B = \sum_{n=0}^{s-1} \frac{1}{n!} \xi^n\zt^{s-1-n} B_n(p_1,p_2,p_3)\,, \\
\label{OTS_ansatz3}
& C = \sum_{n=0}^{s-2} \frac{1}{n!} \xi^n\zt^{s-2-n} C_n(p_1,p_2,p_3)\,.
\end{align}

\subsubsection{Conformal WT identities}

The homogeneity conditions following from the dilatation WT identity are
\begin{align}
&A_n(\lm p_1,\lm p_2,\lm p_3)=\lm^{\Delta_t -2d-s-2} A_n(p_1,p_2,p_3)\,,\label{homo1-02s}\\
&B_n(\lm p_1,\lm p_2,\lm p_3)=\lm^{\Delta_t -2d -s} B_n(p_1,p_2,p_3)\,,\label{homo2-02s}\\
&C_n(\lm p_1,\lm p_2,\lm p_3)=\lm^{\Delta_t -2d-s+2} C_n(p_1,p_2,p_3)\,. \label{homo3-02s}
\end{align}
On the other hand, the special conformal WT identity reads
\begin{align}
0 = (\bsb\cdot\bsK_s + \bsb\cdot\bsK_{\ep_2})\lan \vphi(\bsp_1) \ep_2^2.T(\bsp_2) \ep_3^s.O(\bsp_3) \ran' \,.\label{OTS_CWT}
\end{align}
Just as we did in the previous subsection, we split~\eqref{OTS_CWT} into a set of differential equations just as before using the formula~\eqref{K-Pi} and the identities,
\begin{align}
(\bsb\cdot\bsK_s)A &= (\bsb\cdot\bsp_1)P_1^{(-2,0)}A+(\bsb\cdot\bsp_2)P_2^{(-2,0)}A+(\bsb\cdot\bsep_3)R^{(-2,0)}A
\,,
\\
(\bsb\cdot\bsK_s)B &= (\bsb\cdot\bsp_1)P_1^{(0,1)}B+(\bsb\cdot\bsp_2)P_2^{(0,1)}B+(\bsb\cdot\bsep_3)R^{(0,1)}B
\,,
\\
(\bsb\cdot\bsK_s)C &= (\bsb\cdot\bsp_1)P_1^{(2,2)}C+(\bsb\cdot\bsp_2)P_2^{(2,2)}C+(\bsb\cdot\bsep_3)R^{(2,2)}C
\,.
\end{align}
See Eqs.~\eqref{albt-P1},\eqref{albt-P2}, and \eqref{albt-R} for the definitions of the differential operators $P_{1,2}$ and $R$. After straightforward but tedious algebraic calculations, we obtain the following results.

\paragraph{Primary WT identities}

First, the primary WT identities are $6$ identities associated with $\bsb\cdot\bsp_1$ or $\bsb\cdot\bsp_2$,
\begin{align}
(\bsep_2^2\cdot\bsPi_2\cdot\bsp_1^2)(\bsb\cdot\bsp_1) &: \quad
0=\left( P_1^{(-2,0)}+4p_3^{-2}\te_3 \right)A \,, \label{02s-initial1} \\
(\bsep_2^2\cdot\bsPi_2\cdot\bsp_1^2)(\bsb\cdot\bsp_2) &: \quad
0=\left( P_2^{(-2,0)}+4p_3^{-2}\te_3 \right)A \,, \label{02s-initial2} \\
(\bsep_2^2\cdot\bsPi_2\cdot\bsp_1\bsep_3)(\bsb\cdot\bsp_1) &: \quad
0=\left( P_1^{(0,1)}+2p_3^{-2}\te_3 \right)B+4\pd_\xi A \,, \label{02s-initial3} \\
(\bsep_2^2\cdot\bsPi_2\cdot\bsp_1\bsep_3)(\bsb\cdot\bsp_2) &: \quad
0=\left( P_2^{(0,1)}+2p_3^{-2}\te_3 \right)B \,, \label{02s-initial4} \\
(\bsep_2^2\cdot\bsPi_2\cdot\bsep_3^2)(\bsb\cdot\bsp_1) &: \quad
0=P_1^{(2,2)}C +2\pd_\xi B\,, \label{02s-initial5} \\
(\bsep_2^2\cdot\bsPi_2\cdot\bsep_3^2)(\bsb\cdot\bsp_2) &: \quad
0=P_2^{(2,2)}C\,,\label{02s-initial6}
\end{align}
and 3 identities associated with $\bsb\cdot\bsep_3$,
\begin{align}
(\bsep_2^2\cdot\bsPi_2\cdot\bsp_1^2)(\bsb\cdot\bsep_3) &: \quad
0= \left( R^{(-2,0)}+4\pd_\zeta \right)A + 2p_3^{-2}\te_3 B\, ,\label{02s-recursion1}\\
(\bsep_2^2\cdot\bsPi_2\cdot\bsp_1\bsep_3)(\bsb\cdot\bsep_3) &:\quad
0= \left( R^{(0,1)}+2\pd_\xi + 4\pd_\zeta \right)B + 4 p_3^{-2}\te_3 C\,, \label{02s-recursion2}\\
(\bsep_2^2\cdot\bsPi_2\cdot\bsep_3^2)(\bsb\cdot\bsep_3) &:\quad
0= \left( R^{(2,2)}+4\pd_\xi +4\pd_\zeta \right)C\,. \label{02s-recursion3}
\end{align}

\paragraph{Secondary WT identities}

On the other hand, the secondary WT identities associated with $\bsb\cdot\bsep_2$ are the following two:
\begin{align}
(\bsep_2^2\cdot\bsPi_2\cdot\bsb\,\bsep_3)&: \quad 
0=\left[\te_1+(\xi-\zeta)\pd_\xi+\frac{\bsp_1\cdot\bsp_2}{p_2^2}(d-\te_2)+d-\Delta_1\right]B \nn\\*
&\qquad\qquad
+2\left[\frac{\xi}{p_2^2}(d-\te_2)-\frac{\zeta}{p_3^2}\te_3\right]C\,,\label{02s-second1}\\
(\bsep_2^2\cdot\bsPi_2\cdot\bsb\,\bsp_1)&: \quad 
0=\left[\te_1+(\xi-\zeta)\pd_\xi+\frac{\bsp_1\cdot\bsp_2}{p_2^2}(d-\te_2)+d-\Delta_1+1\right] A\nn\\
&\qquad\qquad
+\left[\frac{\xi}{p_2^2}(d-\te_2)-\frac{\zeta}{p_3^2}\te_3\right]B\,. \label{02s-second2}
\end{align}

\subsubsection{Solving the WT identities}

We now solve the reformulated WT identities. We start with the primary WT identities to specify a form of $A_n$, $B_n$, and $C_n$ with three free parameters. Similarly to the previous case, we use the first six primary WT identities~\eqref{02s-initial1}-\eqref{02s-initial6} to determine the initial conditions. We then solve the recursion relations following from the other three~\eqref{02s-recursion1}-\eqref{02s-recursion3} to obtain all $A_n$, $B_n$, and $C_n$. Finally, we use the secondary WT identities to derive two constraints on the three free parameters, leaving a single free parameter.

\subsubsection*{Initial conditions from primary WT identities}

The $\mathcal{O}(\zeta^0)$ terms of Eqs.~\eqref{02s-initial1}-\eqref{02s-initial6} provide differential equations for $A_s$, $B_{s-1}$, and $C_{s-2}$:
\begin{align}
&0=\left[K_1(\nu_1)-K_3(\nu_3)\right]A_s \,, \label{02s-primary1}\\
&0=\left[K_2(\nu_2)-K_3(\nu_3)\right]A_s \,, \label{02s-primary2}\\
&0=\left[K_1(\nu_1)-K_3(\nu_3)\right]B_{s-1}+4A_s\,,
\label{02s-primary3}\\
&0=\left[K_2(\nu_2)-K_3(\nu_3)\right]B_{s-1}\,, \label{02s-primary4}\\
&0=\left[K_1(\nu_1)-K_3(\nu_3)\right]C_{s-2}+2B_{s-1}\,,
\label{02s-primary5}\\
&0=\left[K_2(\nu_2)-K_3(\nu_3)\right]C_{s-2}\,. \label{02s-primary6}
\end{align}
Following the same strategy as the previous subsection, their general solutions consistent with the homogeneity conditions~\eqref{homo1-02s}-\eqref{homo3-02s} are given in terms of triple-$K$ integrals as
\begin{align}
A_s&=\cC_A J_{s+2\{0,0,0\} }\,,\\
B_{s-1}&=2\cC_A J_{s+1\{1,0,0\}}+\cC_B J_{s\{0,0,0\}}\,,\\
C_{s-2}&=\cC_A J_{s\{2,0,0\}}+\cC_B J_{s-1\{1,0,0\}}+\cC_C J_{s-2\{0,0,0\}}\,,
\end{align}
up to three free parameters $\cC_A$, $\cC_B$, and $\cC_C$.

\subsubsection*{Recursion relations from primary WT identities}

To determine the other $A_n$, $B_n$, and $C_n$, we make use of recursion relations obtained from the identities~\eqref{02s-recursion1}-\eqref{02s-recursion3}:
\begin{align}
0=&\left(\Xi-1+n\right)A_{n+1}-(-s+n)(\Delta_3-1+n)A_n+p_3^{-2}\te_3B_n\,.\label{02s-An}\\*
0=&\left(\Xi+1+n\right)B_{n+1}-(-s+1+n)(\Delta_3+n)B_n+2p_3^{-2}\te_3C_n\,,\label{02s-Bn}\\*
0=&\left(\Xi+3+n\right)C_{n+1}-(-s+2+n)(\Delta_3+1+n)C_n\,.\label{02s-Cn}
\end{align}
Their solutions are
\begin{align}
\nonumber
A_n &=\frac{(\Xi+n-1)_{s-n}}{(-s+n)_{s-n}(\De_3-1+n)_{s-n}}A_{s} \\*
&\quad + \sum_{t=0}^{s-1-n}\frac{(\Xi+n-1)_t}{(-s+n)_{t+1}(\De_3-1+n)_{t+1}}p_3^{-2}\te_3B_{n+t}\,,\\
\nonumber
B_n &=\frac{(\Xi+n+1)_{s-1-n}}{(-s+1+n)_{s-1-n}(\De_3+n)_{s-1-n}}B_{s-1} \\*
&\quad+ 2\sum_{t=0}^{s-2-n}\frac{(\Xi+n+1)_t}{(-s+1+n)_{t+1}(\De_3+n)_{t+1}}p_3^{-2}\te_3C_{n+t}\,,\\
C_n &=\frac{(\Xi+n+3)_{s-2-n}}{(-s+2+n)_{s-2-n}(\De_3+1+n)_{s-2-n}}C_{s-2}\,.
\end{align}

\subsubsection*{A closed form without differential operators}
Just as the two scalar case, it may be convenient to find expressions without the differential operators.
Combining Eq.~\eqref{Xi_on_tripleK} with Eq.~\eqref{k3_down}, we can expand $A_n,B_n,C_n$ in triple-$K$ integrals with a fixed $k_3$ index as
\begin{align}
A_n&=\sum_{k_1,k_2\geq0}a_{n\{k_1,k_2\}}J_{n+k_1+k_2+2\{k_1,k_2,n-s\}}\,,
\\
B_n&=\sum_{k_1,k_2\geq0}b_{n\{k_1,k_2\}}J_{n+k_1+k_2+1\{k_1,k_2,n-s+1\}}\,,
\\
C_n&=\sum_{k_1,k_2\geq0}c_{n\{k_1,k_2\}}J_{n+k_1+k_2\{k_1,k_2,n-s+2\}}\,.
\end{align}
Using Eq.~\eqref{Xi_on_tripleK}, we can compute the coefficients $a_{n\{k_1,k_2\}},b_{n\{k_1,k_2\}},c_{n\{k_1,k_2\}}$ algebraically. The result is summarized as follows: 
The coefficients $a_{n\{k_1,k_2\}}$ with $k_1>0$, $b_{n\{k_1,k_2\}}$ with $k_1>1$, and $c_{n\{k_1,k_2\}}$ with $k_1>2$ all vanish.
Nonzero components of $a_n$ and $b_n$ are given in terms of $c_n$ as
\begin{align}
a_{s-n\{0,k\}}&= c_{s-2-n\{2,k\}} \,,
\quad
b_{s-1-n\{0,k\}}= c_{s-2-n\{1,k\}} \,,
\quad
b_{s-1-n\{1,k\}}= 2c_{s-2-n\{2,k\}} \,.
\end{align}
Finally, nonzero components of $c_n$ are\footnote{Note that the second terms of Eqs.~\eqref{c1k_OTS}-\eqref{c2k_OTS} are interpreted as zero for $k=n$. Similarly, the third term of Eq.~\eqref{c2k_OTS} is zero for $k=n,n-1$. }
\begin{align}
c_{s-2-n\{0,k\}}
&=\cC_C\frac{2^{n-k}(\tfrac{\De_1-\De_2-\De_3-s}{2})_{n-k}(2+\frac{d-s-\De_t}{2})_{n-k}}{k!(n-k)!(2\!-\!\Delta_3\!-\!s)_{n-k}}
\,,
\\*
c_{s-2-n\{1,k\}}&=\cC_B\frac{2^{n-k}(1+\tfrac{\De_1-\De_2-\De_3-s}{2})_{n-k}(1+\frac{d-s-\De_t}{2})_{n-k}}{k!(n-k)!(2\!-\!\Delta_3\!-\!s)_{n-k}}
\nonumber
\\
&\quad
-\cC_C\frac{2^{n-k}(1+\tfrac{\De_1-\De_2-\De_3-s}{2})_{n-k-1}(2+\frac{d-s-\De_t}{2})_{n-k-1}}{k!(n-k-1)!(2\!-\!\Delta_3\!-\!s)_{n-k}}\,,
\label{c1k_OTS}
\\
c_{s-2-n\{2,k\}}&=\cC_A\frac{2^{n-k}(2+\tfrac{\De_1-\De_2-\De_3-s}{2})_{n-k}(\frac{d-s-\De_t}{2})_{n-k}}{k!(n-k)!(2\!-\!\Delta_3\!-\!s)_{n-k}}
\nonumber
\\
&\quad
-\cC_B
\frac{2^{n-k-1}(2+\tfrac{\De_1-\De_2-\De_3-s}{2})_{n-k-1}(1+\frac{d-s-\De_t}{2})_{n-k-1}}{k!(n-k-1)!(2\!-\!\Delta_3\!-\!s)_{n-k}}
\nonumber
\\
\label{c2k_OTS}
&\quad
+\cC_C \frac{2^{n-k-2}(2+\tfrac{\De_1-\De_2-\De_3-s}{2})_{n-k-2}(2+\frac{d-s-\De_t}{2})_{n-k-2}}{k!(n-k-2)!(2\!-\!\Delta_3\!-\!s)_{n-k}}
\,,
\end{align}
where $0 \leq k \leq n$. These coefficients satisfy the identities~\eqref{02s-primary1}-\eqref{02s-primary6} out of the primary WT identities as they satisfy the relations among the coefficients~\eqref{abc_nonzero1}-\eqref{abc_nonzero3} that are equivalent to the six identities.

\subsubsection*{Secondary WT identities}

Finally, we use the secondary WT identities \eqref{02s-second1} and \eqref{02s-second2} to provide constraints on $\cC_A$, $\cC_B$, and $\cC_C$. In the zero momentum limit $p_3\rightarrow0$, these two equations are reduced to
\begin{align}
0&=2\cC_A\Bigr[-j_{s+2,\{0,0,0\}}-j_{s+2\{0,-1,0\}}+(s-\Delta_1-1)j_{s+1\{1,0,0\}}\nn\\
&\qquad \qquad
+(s-1)\left(d\,j_{s\{2,0,0\}}+j_{s-1\{2,-1,0\}}\right)\Bigr] \nn\\
&\quad+\cC_B\Bigr[-j_{s+1,\{-1,0,0\}}-j_{s+1\{0,-1,0\}}+(s-\Delta_1-1)j_{s\{0,0,0\}}
\nn\\
&\qquad\qquad
+2(s-1)\left(d\,j_{s-1\{1,0,0\}}+j_{s\{1,-1,0\}}\right)\Bigr]\nn\\
&\quad+(s-1)\cC_C\left[d\,j_{s-2\{0,0,2\}}+j_{s-1\{0,-1,2\}}\right]
\label{TTS_secondary1}
\end{align}
and
\begin{align}
0&=2\cC_A\Bigr[-j_{s+1,\{-1,0,0\}}-j_{s+1\{0,-1,0\}}+(s-\Delta_1+1)j_{s+2\{0,0,0\}}\nn\\
&\qquad\qquad
+d\,j_{s+1\{1,0,0\}}+j_{s+2\{1,-1,0\}}\Bigr] \nn\\
&\quad
+\cC_B\left[d\,j_{s\{0,0,0\}}+j_{s+1\{0,-1,0\}}\right]\,.
\label{TTS_secondary2}
\end{align}
For example, for a scalar operator dual to a 6-dimensional massless scalar field $\Delta_1=5$ and a spinning operator $s=2,\Delta_3=\frac{11}{2}$,
\begin{align}
\cC_B=-\frac{27707}{4016}\cC_A\,, \quad \cC_C=-\frac{29585757}{76657408}\cC_A\,.
\end{align}
Also for a scalar operator dual to a 6-dimesional scalar field with conformal mass $\Delta_1=4$ and a spinning operator with $s=2,\Delta_3=\frac{11}{2}$,
\begin{align}
\cC_B=-\frac{2757}{1424}\cC_A\,, \quad \cC_C=-\frac{2053563}{113920 }\cC_A\,.
\end{align}
We are now left with a single free parameter for $s\geq2$, which agrees with the position space result. Note that for $s=1$ we have an additional constraint $\cC_C=0$ as we mentioned, hence three-point functions vanish.

\section{Extension to correlators with two conserved currents}
\label{sec:TTs}

In this section we work on correlators with two conserved currents and a general tensor. Since there appear three polarization vectors, the special conformal WT identities become somewhat complicated compared to the previous section. However, we demonstrate that the strategy employed there can be carried over to the present problem without any obstruction: {First, we use the initial conditions and recursion relations following from the primary WT identities to determine correlators up to several free parameters. We then impose constraints on these parameters obtained from the secondary WT identities}. For illustration, we focus on three-point functions,
\begin{align}
\lan \ep_1.J(\bsp_1) \ep_2.J(\bsp_2) \ep_3^s.O(\bsp_3) \ran' \,, \label{11s}
\end{align}
of two identical spin $1$ conserved currents and a general tensor in this section. An extension to the energy-momentum tensor is given in Appendix~\ref{app:TTs}.

\subsection{General ansatz}

Following the strategy used in the single conserved current case, let us first provide the following general ansatz\footnote{
Note that we have $D=0$ for $s=1$ since the $D$ term requires two or more $\bsep_3$. It, however, does not affect our argument very much because three-point functions vanish for odd spin $s$ as we explain shortly.}:
\begin{align}
\lan \ep_1.J(\bsp_1) \ep_2.J(\bsp_2) \ep_3^s.O(\bsp_3) \ran' 
&= 
(\bsep_1\cdot\bspi_1\cdot\bsp_2)(\bsep_2\cdot\bspi_2\cdot\bsp_1)A \nn\\
&\quad 
+(\bsep_1\cdot\bspi_1\cdot\bsp_2)(\bsep_2\cdot\bspi_2\cdot\bsep_3)B
+(\bsep_1\cdot\bspi_1\cdot\bsep_3)(\bsep_2\cdot\bspi_2\cdot\bsp_1)C \nn\\
&\quad 
+(\bsep_1\cdot\bspi_1\cdot\bsep_3)(\bsep_2\cdot\bspi_2\cdot\bsep_3)D
+(\bsep_1\cdot\bspi_1\cdot\bspi_2\cdot\bsep_2)E \,,
\label{11s-ansatz}
\end{align}
where $\bspi_i$ is the transverse projector \eqref{pi} for $\bsp_i$, and $\bsep_1\cdot\bspi_1\cdot\bspi_2\cdot\bsep_2= (\epsilon_1)^\mu(\pi_1)_{\mu\nu}(\pi_2)^{\nu\rho}(\epsilon_2)_\rho$. Also we used the conservation law $\partial_\mu J^\mu=0$, which requires that $\bsep_{1,2}$ have to be contracted with the projector $\bspi_{1,2}$. We then expand each term as
\begin{align}
A &= \sum_{n=0}^s\frac{1}{n!}\xi^n\zt^{s-n}A_n(p_1,p_2,p_3) \,,
\label{expansionA}
\\
B &= \sum_{n=0}^{s-1}\frac{1}{n!}\xi^n\zt^{s-1-n}B_n(p_1,p_2,p_3) \,, \\
C &= \sum_{n=0}^{s-1}\frac{1}{n!}\xi^n\zt^{s-1-n}C_n(p_1,p_2,p_3) \,, \\
D &= \sum_{n=0}^{s-2}\frac{1}{n!}\xi^n\zt^{s-2-n}D_n(p_1,p_2,p_3) \,, \\
E &= \sum_{n=0}^s\frac{1}{n!}\xi^n\zt^{s-n}E_n(p_1,p_2,p_3) \,.
\label{expansionE}
\end{align}
Since we are considering two identical conserved currents $J_\mu$, three-point functions are symmetric under the $1 \leftrightarrow 2$ exchange: $A$, $B+C$, $D$, and $E$ carry an even parity under the exchange, whereas $B-C$ has an odd parity. In particular, it requires the following relations among the initial conditions,
\begin{align}
A^{}_s(p_1,p_2,p_3)&=(-1)^sA_s(p_2,p_1,p_3) \,, \label{11s-flip1} \\
B_{s-1}(p_1,p_2,p_3)&=(-1)^{s-1}C_{s-1}(p_2,p_1,p_3) \,, \label{11s-flip2} \\
D_{s-2}(p_1,p_2,p_3)&=(-1)^{s-2}D_{s-2}(p_2,p_1,p_3) \,, \label{11s-flip3} \\
E_s(p_1,p_2,p_3)&=(-1)^sE_s(p_2,p_1,p_3) \,. \label{11s-flip4} 
\end{align}
Similar relations hold for other terms in the expansion~\eqref{expansionA}-\eqref{expansionE}.

\subsection{Reformulating WT identities}

We next reformulate the conformal WT identities. In terms of Eqs.~\eqref{expansionA}-\eqref{expansionE}, the dilatation WT identity yields the following homogeneity conditions:
\begin{align}
A_n(\lm p_1,\lm p_2,\lm p_3) &= \lm^{\De_t-2d-s-2} A_n(p_1,p_2,p_3) \,, 
\label{JJS_d1}\\
B_n(\lm p_1,\lm p_2,\lm p_3) &= \lm^{\De_t-2d-s} B_n(p_1,p_2,p_3) \,, \\
C_n(\lm p_1,\lm p_2,\lm p_3) &= \lm^{\De_t-2d-s} C_n(p_1,p_2,p_3) \,, \\
D_n(\lm p_1,\lm p_2,\lm p_3) &= \lm^{\De_t-2d-s+2} D_n(p_1,p_2,p_3) \,, \\
E_n(\lm p_1,\lm p_2,\lm p_3) &= \lm^{\De_t-2d-s} E_n(p_1,p_2,p_3) \,.
\label{JJS_d5}
\end{align}
On the other hand, to rewrite the special conformal WT identity,
\begin{align}
0 = (\bsb\cdot\bsK_s + \bsb\cdot\bsK_{\ep_1} + \bsb\cdot\bsK_{\ep_2})\lan \ep_1.J(\bsp_1) \ep_2.J(\bsp_2) \ep_3^s.O(\bsp_3) \ran' \,,\label{JJS_CWT}
\end{align}
it is convenient to note
\begin{align}
(\bsb\cdot\bsK_s)A &= (\bsb\cdot\bsp_1)P_1^{(-2,0)}A+(\bsb\cdot\bsp_2)P_2^{(-2,0)}A+(\bsb\cdot\bsep_3)R^{(-2,0)}A
\,,
\\
(\bsb\cdot\bsK_s)B &= (\bsb\cdot\bsp_1)P_1^{(0,1)}B+(\bsb\cdot\bsp_2)P_2^{(0,1)}B+(\bsb\cdot\bsep_3)R^{(0,1)}B
\,,
\\
(\bsb\cdot\bsK_s)C &= (\bsb\cdot\bsp_1)P_1^{(0,1)}C+(\bsb\cdot\bsp_2)P_2^{(0,1)}C+(\bsb\cdot\bsep_3)R^{(0,1)}C
\,,
\\
(\bsb\cdot\bsK_s)D &= (\bsb\cdot\bsp_1)P_1^{(2,2)}D+(\bsb\cdot\bsp_2)P_2^{(2,2)}D+(\bsb\cdot\bsep_3)R^{(2,2)}D
\,,
\\
(\bsb\cdot\bsK_s)E &= (\bsb\cdot\bsp_1)P_1^{(0,0)}E+(\bsb\cdot\bsp_2)P_2^{(0,0)}E+(\bsb\cdot\bsep_3)R^{(0,0)}E
\,.
\end{align}
Using these formulae and Eq.~\eqref{K_pi}, after a straightforward but lengthy calculation, we arrive at the following primary and secondary WT identities.

\subsubsection*{Primary WT identities}

First, the primary WT identities consist of 15 equations, 10 of which are associated with $\bsb\cdot\bsp_i$ ($i=1,2$) of the form,
\begin{align}
0&=\left(P_1^{(-2,0)}+4p_3^{-2}\te_3\right)A \,, \label{11s-initial1} \\
0&=\left(P_2^{(-2,0)}+4p_3^{-2}\te_3\right)A \,, \label{11s-initial2} \\
0&=\left(P_1^{(0,1)}+2p_3^{-2}\te_3\right)B+2\pd_\xi A \,, \label{11s-initial3} \\*
0&=\left(P_2^{(0,1)}+2p_3^{-2}\te_3\right)B \,, \label{11s-initial4} \\
0&=\left(P_1^{(0,1)}+2p_3^{-2}\te_3\right)C \,, \label{11s-initial5} \\
0&=\left(P_2^{(0,1)}+2p_3^{-2}\te_3\right)C-2\pd_\xi A \,, \label{11s-initial6} \\
0&=P_1^{(2,2)}D+2\pd_\xi C \,, \label{11s-initial7} \\
0&=P_2^{(2,2)}D-2\pd_\xi B \,, \label{11s-initial8} \\
0&=P_1^{(0,0)}E+2A \,, \label{11s-initial9} \\
0&=P_2^{(0,0)}E+2A \,. \label{11s-initial10} 
\end{align}
We will use them to determine the initial conditions  $A_s$, $B_{s-1}$, $C_{s-1}$, $D_{s-2}$, and $E_s$. The other 5 are associated to $\bsb\cdot\bsep_3$ and given by
\begin{align}
0&=\left(R^{(-2,0)}+2\pd_\xi+4\pd_\zt\right)A+2p_3^{-2}\te_3(B+C) \,, \label{11s-rr1} \\
0&=\left(R^{(0,1)}+4\pd_\xi+4\pd_\zt\right)B+2p_3^{-2}\te_3D \,, \label{11s-rr2} \\
0&=\left(R^{(0,1)}+4\pd_\zt\right)C+2p_3^{-2}\te_3D \,, \label{11s-rr3} \\
0&=\left(R^{(2,2)}+2\pd_\xi+4\pd_\zt\right)D \,, \label{11s-rr4} \\
0&=R^{(0,0)}E+2(B+C) \,, \label{11s-rr5} 
\end{align}
which provide recursion relations among $A_n,...,E_n$. These primary WT identities fix the functional form of three-point functions up to five free parameters.

\subsubsection*{Secondary WT identities}

On the other hand, the secondary WT identities consist of 4 equations associated with $\bsb\boldsymbol{\cdot\pi_i\cdot}\bsep_i$ ($i=1,2$) of the form,
\begin{align}
0&=\left[ \te_2+\xi\pd_\xi+d-\De_2+\frac{\bsp_1\cdot\bsp_2}{p_1^2}(d-2-\te_1) \right]A \nn\\
&\quad\quad\quad 
+\left[ \frac{\zt-\xi}{p_1^2}(d-2-\te_1)-\frac{\zt}{p_3^2}\te_3 \right]C
+\frac{1}{p_1^2}(d-2-\te_1)E \,, \label{11s-sec1} \\
0&=\left[ \te_1+(\xi-\zt)\pd_\xi+d-\De_1+\frac{\bsp_1\cdot\bsp_2}{p_2^2}(d-2-\te_2) \right]A \nn\\*
&\quad\quad\quad 
+\left[ \frac{\xi}{p_2^2}(d-2-\te_2)-\frac{\zt}{p_3^2}\te_3 \right]B
+\frac{1}{p_2^2}(d-2-\te_2)E \,, \label{11s-sec2} \\
0&=\left[ \te_2+\xi\pd_\xi+d-\De_2+\frac{\bsp_1\cdot\bsp_2}{p_1^2}(d-2-\te_1) \right]B
-C \nn\\*
&\quad\quad\quad 
+\left[ \frac{\zt-\xi}{p_1^2}(d-2-\te_1)-\frac{\zt}{p_3^2}\te_3 \right]D
+\pd_\xi E \,, \label{11s-sec3} \\
0&=\left[ \te_1+(\xi-\zt)\pd_\xi+d-\De_1+\frac{\bsp_1\cdot\bsp_2}{p_2^2}(d-2-\te_2) \right]C
-B
\nonumber
\\*
&\quad
+\left[ \frac{\xi}{p_2^2}(d-2-\te_2)-\frac{\zt}{p_3^2}\te_3 \right]D
-\pd_\xi E \,. \label{11s-sec4} 
\end{align}
Note that symmetry under the exchange $1\leftrightarrow2$ implies that the first two equations are equivalent, and the third and the fourth also. We therefore have only two independent equations to be considered once the exchange symmetry is taken into account.

\subsection{Solving WT identities}

We proceed to solving the reformulated WT identities. Just as in the single conserved current case, we use the primary WT identities to determine the form of three-point functions up to several free parameters and then use the secondary to provide constraints on them.

\subsubsection*{Initial conditions from primary WT identities}

The 10 identities \eqref{11s-initial1}-\eqref{11s-initial10} provide a set of recursion relations for $A_n,...,E_n$, which are somewhat complicated to solve for general $n$. However, their $\mathcal{O}(\zeta^0)$ parts provide equations for $A_s$, $B_{s-1}$, $C_{s-1}$, $D_{s-2}$, and $E_s$ only:
\begin{align}
0&=[K_1(\nu_1)-K_3(\nu_3)]A_s \,, \label{11s-primary1} \\
0&=[K_2(\nu_2)-K_3(\nu_3)]A_s \,, \label{11s-primary2} \\
0&=[K_1(\nu_1)-K_3(\nu_3)]B_{s-1}+2A_s \,,
\label{11s-primary3} \\
0&=[K_2(\nu_2)-K_3(\nu_3)]B_{s-1} \,, \label{11s-primary4} \\
0&=[K_1(\nu_1)-K_3(\nu_3)]C_{s-1} \,, \label{11s-primary5} \\
0&=[K_2(\nu_2)-K_3(\nu_3)]C_{s-1}-2A_s \,,
\label{11s-primary6} \\
0&=[K_1(\nu_1)-K_3(\nu_3)]D_{s-2}+2C_{s-1} \,,
\label{11s-primary7} \\
0&=[K_2(\nu_2)-K_3(\nu_3)]D_{s-2}-2B_{s-1} \,,
\label{11s-primary8} \\
0&=[K_1(\nu_1)-K_3(\nu_3)]E_s+2A_s \,, \label{11s-primary9} \\
0&=[K_2(\nu_2)-K_3(\nu_3)]E_s+2A_s \,. \label{11s-primary10} 
\end{align}
Their general solutions consistent with the homogeneity conditions~\eqref{JJS_d1}-\eqref{JJS_d5} can be constructed in the same manner as the previous section as
\begin{align}
A_s(p_1,p_2,p_3) &= \cC_A J_{s+2(0,0,0)}(p_1,p_2,p_3) \,.\label{11s-A}
\\
B_{s-1}(p_1,p_2,p_3) &= \cC_B J_{s(0,0,0)}(p_1,p_2,p_3)+\cC_AJ_{s+1(1,0,0)}(p_1,p_2,p_3)\,,
\\
C_{s-1}(p_1,p_2,p_3) &= \cC_C J_{s(0,0,0)}(p_1,p_2,p_3)-\cC_AJ_{s+1(0,1,0)}(p_1,p_2,p_3)\,,
\\
D_{s-2}(p_1,p_2,p_3) &= \cC_D J_{s-2(0,0,0)}(p_1,p_2,p_3)-\cC_AJ_{s(1,1,0)}(p_1,p_2,p_3)
\nonumber\\*
&\quad
-\cC_B J_{s-1(0,1,0)}(p_1,p_2,p_3)+\cC_C J_{s-1(1,0,0)}(p_1,p_2,p_3)
\,,
\\
E_s(p_1,p_2,p_3) &= \cC_E J_{s(0,0,0)}(p_1,p_2,p_3)-\cC_AJ_{s+1(0,0,1)}(p_1,p_2,p_3)
\end{align}
with five free parameters $\cC_A$, $\cC_B$, $\cC_C$, $\cC_D$, and $\cC_E$.

\subsubsection*{Recursion relations}

Next we solve the other five primary WT identities \eqref{11s-rr1}-\eqref{11s-rr5}, which provide the following recursion relations:
\begin{align}
0=&\left(\Xi+n\right)A_{n+1}-(-s+n)(\Delta_3-1+n)A_n+p_3^{-2}\te_3(B_n+C_n)\,,\label{11s-re1}\\
0=&\left(\Xi+2+n\right)B_{n+1}-(-s+1+n)(\Delta_3+n)B_n+p_3^{-2}\te_3D_n\,,\label{11s-re2}\\
0=&\left(\Xi+n\right)C_{n+1}-(-s+1+n)(\Delta_3+n)C_n+p_3^{-2}\te_3D_n\,,\label{11s-re3}\\
0=&\left(\Xi+2+n\right)D_{n+1}-(-s+2+n)(\Delta_3+1+n)D_n\,,\label{11s-re4}\\
0=&\left(\Xi+n\right)E_{n+1}-(-s+n)(\Delta_3-1+n)E_n+p_3^{-2}\te_3(B_n+C_n)\,.\label{11s-re5}
\end{align}
Their solutions are
\begin{align}
A_n &= \frac{(\Xi+n)_{s-n}}{(-s+n)_{s-n}(\De_3-1+n)_{s-n}}A_s
\nonumber
\\*
&\quad
+\sum_{t=0}^{s-1-n}\frac{(\Xi+n)_t}{(-s+n)_{t+1}(\De_3-1+n)_{t+1}}p_3^{-2}\te_3(B_{n+t}+C_{n+t})\,,\\
B_n &= \frac{(\Xi+n+2)_{s-1-n}}{(-s+1+n)_{s-1-n}(\De_3+n)_{s-1-n}}B_{s-1}
\nonumber
\\*
&\quad
+\sum_{t=0}^{s-2-n}\frac{(\Xi+n+2)_t}{(-s+1+n)_{t+1}(\De_3+n)_{t+1}}p_3^{-2}\te_3D_{n+t}\,,\\
C_n &= \frac{(\Xi+n)_{s-1-n}}{(-s+1+n)_{s-1-n}(\De_3+n)_{s-1-n}}C_{s-1}
\nonumber
\\*
&\quad
+ \sum_{t=0}^{s-2-n}\frac{(\Xi+n)_t}{(-s+n+1)_{t+1}(\De_3+n)_{t+1}}p_3^{-2}\te_3D_{n+t}\,,\\
D_n &= \frac{(\Xi+n+2)_{s-2-n}}{(-s+2+n)_{s-2-n}(\De_3+1+n)_{s-2-n}}D_{s-2}\,,\\
E_n &= \frac{(\Xi+n)_{s-n}}{(-s+n)_{s-n}(\De_3-1+n)_{s-n}}E_s
\nonumber
\\*
&\quad
+ \sum_{t=0}^{s-1-n}\frac{(\Xi+n)_t}{(-s+n)_{t+1}(\De_3-1+n)_{t+1}}p_3^{-2}\te_3(B_{n+t}+C_{n+t})\,.
\end{align}

\subsubsection*{Constraints from $1\leftrightarrow2$ exchange symmetry}

Before moving on to the secondary WT identities, let us consider implications from the $1\leftrightarrow2$ exchange symmetry of the two conserved currents. The four conditions~\eqref{11s-flip1}-\eqref{11s-flip4} are now translated into the constraints on the five free parameters as
\begin{align}
\text{even $s$:}&\quad \cC_B=-\cC_C\,,
\\*
\text{odd $s$:}&\quad \cC_B=\cC_C\,,
\quad\cC_A=\cC_D=\cC_E=0\,.
\end{align}
Therefore, there remain four free parameters for even spin $s$, whereas a single parameter for odd spin $s$. Note that the full correlator is consistent with the exchange symmetry once we impose its constraint on the initial conditions. It is because the WT identities used to derive the recursion relations are compatible with the exchange symmetry.

\subsubsection*{Secondary WT identities}

Finally, let us impose the secondary WT identities. As we mentioned, there remain two independent secondary WT identities once we require the $1\leftrightarrow2$ exchange symmetry. An immediate observation is that three-point functions vanish for odd spin $s$ because there exist two constraints on a single parameter\footnote{To be precise, we need to check that the two secondary WT identities provide nontrivial conditions on $\cC_B$. However, it is easy to show that it is indeed the case and thus $\cC_A=\cC_B=\cC_C=\cC_D=\cC_E=0$ is required for odd spin $s$ by the exchange symmetry and the secondary WT identities.}. On the other hand, for even spin $s$, there are four free parameters $\cC_A$, $\cC_B=-\cC_C$, $\cC_D$, and $\cC_E$ after imposing the exchange symmetry. The secondary WT identities then provide two independent constraints on the four parameters, leaving two free parameters afterwards. A general form of the two conditions is somewhat complicated, but it is straightforward to derive these two in the same manner as the previous section. For example, for a spinning operator with $\Delta_3=\frac{11}{2},s=2$ in 5 dimension, we find
\begin{align}
\cC_B=-\frac{1}{2}\cC_A-\frac{1}{2}\cC_E\,,
\quad
\cC_D=\frac{291}{128}\cC_A+\frac{611}{192}\cC_E\,.
\end{align}

\bigskip
To summarize, we have provided an expression for three-point functions in terms of triple-$K$ integrals and differential operators. Just as the two scalar and one scalar cases, we may use Eqs.~\eqref{Euler_on_tripleK}, \eqref{k3_down}, and \eqref{Xi_on_tripleK} to rewrite $A_n$, $B_n$, $C_n$, $D_n$, and $E_n$ in terms of triple-$K$ integrals into a form similar to the expansion~\eqref{tripleK_expansion}. The coefficients can be found algebraically, e.g., with the help of computer software, even though we leave derivation of a closed form for general cases for future work. Also, we have not explicitly shown that our expression for the full correlator satisfies Eqs.~\eqref{11s-initial1}-\eqref{11s-initial10} and~\eqref{11s-sec1}-\eqref{11s-sec4}. Since it has the correct number of free parameters (known in position space~\cite{Costa:2011mg}), it should satisfy them automatically. We have checked that it is indeed the case for several examples, leaving a general proof for future work. The same remark applies to correlators with two energy-momentum tensors studied in Appendix~\ref{app:TTs}.

\section{Summary and outlook}
\setcounter{equation}{0}

In this paper we constructed conformal three-point functions with a symmetric traceless tensor and conserved currents in momentum space\footnote{
Our strategy will be applicable to correlators with an antisymmetric tensor or a more general tensor with mixed symmetry by introducing Grassmann odd polarization vectors, which we leave for future work.}. Reformulating the conformal WT identities into the primary and secondary ones, we decomposed our problem into the following two steps: We first used the primary WT identities to derive recursion relations among functional coefficients in the tensor decomposition and determine the initial conditions. This step specifies three-point functions up to several free parameters. We then used the secondary WT identities to provide constraints on these parameters. Our expression is based on triple-$K$ integrals and a differential operator which relates triple-$K$ integrals with different indices. 
For correlators with no or one conserved current, we found explicit forms without the differential operator based on the expansion by triple-$K$ integrals.

\medskip
We would like to conclude the paper with several future directions. First, our present work will be useful for the study of four and higher point functions with conserved currents. In~\cite{Isono:2018rrb}, based on symmetries and analyticity, a crossing symmetric basis of scalar four-point functions with a general intermediate operator was constructed. There, three-point functions of two scalars and a general tensor was useful because general tensors appear as intermediate states. It would be interesting to extend the construction to four-point functions with conserved currents. We expect that such a direction will be useful, e.g., for the Polyakov type bootstrap approach~\cite{Polyakov:1974gs, Sen:2015doa, Gopakumar:2016wkt, Gopakumar:2016cpb,Gopakumar:2018xqi}. Another promising direction is cosmological applications. In~\cite{Arkani-Hamed:2018kmz}, in the same spirit as~\cite{Isono:2018rrb}, a basis of de Sitter four-point functions of a massless scalar was constructed and its cosmological implication was discussed. It would be interesting to construct a similar basis for primordial graviton non-Gaussianities extending our present work. We hope to report our progress in these directions elsewhere.

\section*{Acknowledgements}
H.I. is supported in part by the ``CUniverse'' research promotion project by Chulalongkorn University (grant reference CUAASC).
T.N. is supported in part by JSPS KAKENHI Grant Numbers JP17H02894 and JP18K13539, and MEXT KAKENHI Grant Number JP18H04352.


\appendix

\section{Properties of the differential operators $\bsK_s$ and $\bsK_{\epsilon_i}$}
\setcounter{equation}{0}
\label{appsec:Ks}

We summarize properties of the differential operators $\bsK_s$ and $\bsK_{\epsilon_i}$ appearing in the special conformal WT identity~\eqref{eq:SCWT}.

\subsection{Rewriting $\bsK_s$}

Let us first consider $\bsb\cdot\bsK_s$ acting on the following function with free parameters $\al$ and $\bt$:
\begin{align}
F = \sum_{n=0}^{s-\bt}\frac{1}{n!} \xi^n \zt^{s-\bt-n} f_n(p_1,p_2,p_3)\,,
\end{align}
where $\xi=\bsep_3\cdot\bsp_2$ and $\zt=\bsep_3\cdot(\bsp_1+\bsp_2)$ with the null helicity vector $\bsep_3$, and the function $f_n$ is homogeneous of degree $\De_t-2d-s+\al$,
\begin{align}
f_n(\lm p_1,\lm p_2,\lm p_3) = \lm^{\De_t-2d-s+\al}f_n(p_1,p_2,p_3)\,.
\end{align}
Recall that $p_3$ should be understood as $p_3=|\bsp_1+\bsp_2|$.
Let us rewrite $(\bsb\cdot\bsK_s) F$ as a linear combination of $\bsb\cdot\bsp_1$, $\bsb\cdot\bsp_2$, and $\bsb\cdot\bsep_3$:
\begin{align}
(\bsb\cdot\bsK_s) F = (\bsb\cdot\bsp_1)P_1^{(\alpha,\beta)}F+(\bsb\cdot\bsp_2)P_2^{(\alpha,\beta)}F+(\bsb\cdot\bsep_3)R^{(\alpha,\beta)}F \, ,
\end{align}
where $P_1^{(\alpha,\beta)}$, $P_2^{(\alpha,\beta)}$, and $R^{(\alpha,\beta)}$ are differential operators given shortly.
The equation $0=(\bsb\cdot\bsK_s) F$ is then equivalent to 
\begin{align}
P_1^{(\alpha,\beta)}F=P_2^{(\alpha,\beta)}F=R^{(\alpha,\beta)}F=0 \, .
\end{align}
In terms of the differential operators with respect to $p_1$, $p_2$, $p_3$, $\xi$, and $\zt$,
the coefficients $P_1^{(\alpha,\beta)}F$ and $P_2^{(\alpha,\beta)}F$ are given by
\begin{align}
P_1^{(\alpha,\beta)}F &= [\cK_1(\nu_1) - \cK_3(\nu_3) + 2(\al-\bt-\zt\pd_\zt)p_3^{-2}\te_{3}] ~ F \,, \label{albt-P1} \\
P_2^{(\alpha,\beta)}F &= [\cK_2(\nu_2) - \cK_3(\nu_3) + 2(\al-\bt-\zt(\pd_\xi+\pd_\zt))p_3^{-2}\te_{3}] ~ F \,, \label{albt-P2}
\end{align}
where the differential operator $\cK_i(\nu_i)$ is defined by
\begin{align}
\cK_i(\nu_i) = p_i^{-2}\te_{i}(\te_{i}-2\nu_i) \label{K_i}\, .
\end{align}
On the other hand, $R^{(\alpha,\beta)}F$ may be expressed with the new variable $x=\xi/\zt$ as
\begin{align}
\label{albt-R}
R^{(\alpha,\beta)}F &= 2x^{-1}\zt^{s-\bt-1}\left[\te_x\left(\te_x+\Xi+\frac{\al}{2}-1\right) - x(\te_x+\bt-s)(\te_x+\De_3-1+\al-\bt)\right] \nn\\
&\qquad \times \sum_{n=0}^{s-\bt} \frac{x^n}{n!} f_n\,,
\end{align}
where the differential operator $\Xi$ is defined as
\begin{align}
\Xi = \frac{1}{2}\bigg( \De_1-\De_2+\De_3-s-\te_{1}+\te_{2}-\frac{p_1^2-p_2^2}{p_3^2}\te_{3} \bigg)\,.
\end{align}
Note that when deriving these formulae, we used the homogeneity condition,
\begin{align}
(\te_{1}+\te_{2}+\te_{3})f_n=(\De_t-2d-s+\al)f_n \,.
\end{align}
It is also convenient to rewrite the expression \eqref{albt-R} as
\begin{align}
\label{albt-R-rec}
R^{(\alpha,\beta)}F=2\zt^{s-\bt-1} \sum_{n=0}^{s-\bt} \frac{x^n}{n!} \left[
(\Xi+\tfrac{\al}{2}+n)f_{n+1} - (\bt-s+n)(\De_3-1+\al-\bt+n)f_n
\right]
\end{align}
with $f_{s-\bt+1}=0$. We use it to derive recursion relations among $f_n$.

\subsection{Formulae for $\bsK_s+\bsK_{\epsilon_i}$}

In the special conformal WT identity for correlators with conserved currents, the differential operator $\bsb\cdot\bsK_s+\bsb\cdot\bsK_{\epsilon_i}$ acts on the projectors $(\pi_i)_{\mu\nu}$ and $(\Pi_i)_{\mu\nu\rho\sig}$.
To massage the complicated identities, it is convenient to find identities about the commutation relation of $\bsb\cdot\bsK_s+\bsb\cdot\bsK_{\epsilon_i}$ and the projectors.

\medskip
Let us first consider correlators with a spin $1$ conserved current with momentum $\bsp_i$. The correlator has the tensor structure $(\bsep_i\cdot\bspi_i)_{\mu}X^{\mu}$, where $X^{\mu}$ is an arbitrary vector function without $\bsep_i$ dependence. We also used the shorthand notation $(\bsep_i\cdot\bspi_i)_{\mu}=(\ep_i)^\nu(\pi_i)_{\nu\mu}$. Then, the action of $\bsb\cdot\bsK_s+\bsb\cdot\bsK_{\epsilon_i}$ reads
\begin{align}
 &({\bsb\cdot\bsK_s}+{\bsb\cdot\bsK_{\epsilon_i}})({\bsep_i\cdot\bspi_i})_{\mu}X^\mu\nonumber\\*
 &~~ = (\bsep_i\cdot\bspi_i)_{\mu}({\bsb\cdot\bsK_s}) X^{\mu}\nonumber \\*
 &~~~~ +2\left({\bsep_{i}\cdot\bspi_i\cdot{\bspd_{i}}}\right)(\bsb \cdot \bsX)
 -2({\bsb\cdot\bspi_i\cdot\bsep_{i}})(\bspd_i\cdot \bsX)+\frac{2(d-2)}{p_i^2}({\bsb\cdot\bspi_i\cdot\bsep_{i}})(\bsp_2 \cdot \bsX) \,, \label{K_pi}
\end{align}
where $\bspd_i=\pd/\pd\bsp_i$.

\medskip
Next we turn to correlators with the energy-momentum tensor. The correlator has the tensor structure $({\bsep_i^2\cdot\bsPi_i})_{\mu\nu}X^{\mu\nu}$, where $X^{\mu\nu}$ is an arbitrary tensorial function without $\bsep_i$ dependence. Also we used the notation $({\bsep_i^2\cdot\bsPi_i})_{\mu\nu}=(\ep_i)^\rho(\ep_i)^\sig(\Pi_i)_{\rho\sig\mu\nu}$. We then find
\begin{align}
 &({\bsb\cdot\bsK_s}+{\bsb\cdot\bsK_{\epsilon_i}})({\bsep_i^2\cdot\bsPi_i})_{\mu\nu}X^{\mu\nu}\nonumber\\
 &=({\bsep_i^2\cdot\bsPi_i})_{\mu\nu}({\bsb\cdot\bsK_s})X^{\mu\nu} \nonumber\\
 &\quad
 +2\left[({\bsep_{i}^2\cdot\bsPi_i\cdot{\bspd_{i}}})_{\mu}b_{\nu} - ({\bsep_i^2\cdot\bsPi_i\cdot \bsb})_\mu(\pd_i)_\nu
 +d(\bsep_i^2\cdot\bsPi_i\cdot \bsb)_\mu (p_i)_\nu\right]\label{K-Pi}(X^{\mu\nu}+X^{\nu\mu})\,,
\end{align}
where $(\pd_i)_\nu=\pd/\pd (p_i)^\nu$ and $(\bsep_i^2\cdot\bsPi_i\cdot \bsa)_\mu=(\ep_i)^\rho(\ep_i)^\sig(\Pi_i)_{\rho\sig\nu\mu}a^\nu$.

\bigskip
Here let us recall that the conservation law is compatible with conformal symmetry only when $\Delta=d-2+s$ without anomalous dimension. Indeed, we have used this relation to derive Eqs.~\eqref{K_pi} and \eqref{K-Pi}, where the helicity vector $\bsep_i$ is coupled to projectors on both sides consistently. 

\section{Triple-$K$ integrals}
\label{app:tripleK}

In this appendix we summarize various properties of triple-$K$ integrals.

\subsection{Definition}

Let us begin with the special conformal WT identities,
\begin{align}
0&= [\cK_1(\nu_1) - \cK_3(\nu_3)] ~ F(p_1,p_2,p_3) \,,
\label{triple-K1} \\
0&= [\cK_2(\nu_2) - \cK_3(\nu_3)] ~ F(p_1,p_2,p_3) \,, \label{triple-K2}
\end{align}
and the dilatation WT identity,
\begin{align}
F(\lambda p_1,\lambda p_2,\lambda p_3)=\lambda^{\De_t-2d}F(p_1,p_2,p_3)\,,
\label{triple-K3}
\end{align}
of scalar three-point functions, where $\cK_i(\nu_i)=p_i^{-2}\te_{i}(\te_{i}-2\nu_i)$.
If we require that there is no singularity in the domain $p_i>0$, their solution can uniquely be determined up to an overall constant as~\cite{Bzowski:2013sza}
\begin{align}
\label{triple-K_simple}
F(p_1,p_2,p_3)=\int_{0}^{\infty}\frac{dz}{z}z^{\Delta_t-2d}\prod_{i=1}^3(p_iz)^{\nu_i}K_{\nu_i}(p_iz)\,.
\end{align}
Here $K_{\nu}(x)$ is the modified Bessel function of the second kind, which we call the Bessel $K$ function. It is defined by a hypergeometric series\footnote{The Bessel $K$ function $K_n(x)$ with an integer index $n$ is defined by the limit $\displaystyle K_n(x)=\lim_{\epsilon\rightarrow 0} K_{n+\epsilon}(x)$.},
\begin{align}
K_\nu(x)=\frac{\pi}{2\sin(\pi\nu)}\left[I_{-\nu}(x)-I_{\nu}(x)\right]
\quad
{\rm with}
\quad
I_\nu(x)=\sum_{j=0}^{\infty}\frac{1}{j!\Gamma (\nu +j+1)}\left(\frac{x}{2}\right)^{\nu +2j}\,,
\end{align}
and satisfies Bessel's equation,
\begin{align}
\left(\theta_x^2-\nu^2\right)K_\nu(x)=x^2K_\nu(x)\,.
\end{align}
Note that the integral~\eqref{triple-K_simple} is convergent only when $|{\rm Re}\,\nu_1|+|{\rm Re}\,\nu_2|+|{\rm Re}\,\nu_3|<\frac{d}{2}$. Otherwise, there appears a singularity near $z=0$ and we need to perform analytic continuation~\cite{Bzowski:2015pba,Bzowski:2015yxv}, which may be carried out, e.g., by introducing the Pochhammer contour.

\medskip
To construct three-point functions with tensors, it is convenient to generalize the integral~\eqref{triple-K_simple} to the following triple-$K$ integral with indices $N$ and $k_i$ ($i=1,2,3$):
\begin{align}
J_{N\{k_1,k_2,k_3\}}(p_{1},p_{2},p_{3})
 & =\int_{0}^{\infty}\frac{dz}{z}\:z^{2d-\Delta_{t}-k_t+N}
 \prod_{i=1}^3\left(p_{i}z\right)^{\nu_{i}+k_i}K_{\nu_{i}+k_i}(p_{i}z)
 \,,
\end{align}
where $k_t=k_1+k_2+k_3$. It satisfies the differential equations,
\begin{align}
0&= [\cK_1(\nu_1+k_1) - \cK_3(\nu_3+k_3)] ~ J_{N\{k_1,k_2,k_3\}}(p_{1},p_{2},p_{3}) \,,
\\
0&= [\cK_2(\nu_2+k_2) - \cK_3(\nu_3+k_3)] ~ J_{N\{k_1,k_2,k_3\}}(p_{1},p_{2},p_{3}) \,,
\end{align}
and the homogeneity conditions,
\begin{align}
J_{N\{k_1,k_2,k_3\}}(\lambda p_1,\lambda p_2,\lambda p_3)=\lambda^{\De_t+k_t-N-2d}J_{N\{k_1,k_2,k_3\}}(p_{1},p_{2},p_{3})\,.
\end{align}

\subsection{Differential operators acting on triple-$K$ integrals}

We then demonstrate how various differential operators act on triple-$K$ integrals, which is useful when we solve the WT identities in the main text. The origin of all the formulae below is the following action of the Euler operator on the Bessel $K$ function:
\begin{align}
\theta_x\big(x^\nu K_\nu(x)\big)=-x^2\big(x^{\nu-1} K_{\nu-1}(x)\big)=
2\nu x^\nu K_\nu(x)-x^{\nu+1}K_{\nu+1}(x)\,.
\end{align}
This can be translated into
\begin{align}
\label{Euler_on_tripleK}
\te_1J_{N\{k_1,k_2,k_3\}}=-p_1^2J_{N+1\{k_1-1,k_2,k_3\}}
=2(\nu_1+k_1)J_{N\{k_1,k_2,k_3\}} -J_{N+1\{k_1+1,k_2,k_3\}}
\,.
\end{align}
Here and in what follows we occasionally omit explicit indication of momentum dependence. It then follows that
\begin{align}
\label{K_on_tripleK}
\cK_1(\nu_1)J_{N\{k_1,k_2,k_3\}}=-2k_{1}J_{N+1\{k_{1}-1,k_{2},k_{3}\}}+J_{N+2\{k_{1},k_{2},k_{3}\}}\,.
\end{align}
Similar relations hold for $\theta_{2,3}$ and $\cK_{2,3}$. Combining Eq.~\eqref{Euler_on_tripleK} with the homogeneity condition of the triple-$K$ integral,
\begin{align}
(\te_1+\te_2+\te_3)J_{N\{k_1,k_2,k_3\}}(p_1,p_2,p_3)&=\lambda\frac{d}{d\,\lambda}J_{N\{k_1,k_2,k_3\}}(\lambda p_1,\lambda p_2,\lambda p_3)\Big|_{\lambda=1} \nn\\
&=(\Delta_t+k_t-N-2d)J_{N\{k_1,k_2,k_3\}}(p_1,p_2,p_3)\,,\label{homo-K}
\end{align}
we also find identities among the nearest neighbors,
\begin{align}
\label{k3_down}
&(\Delta_t+k_t+N-d)J_{N\{k_1,k_2,k_3\}} \nonumber\\
&\qquad=J_{N+1\{k_1+1,k_2,k_3\}}+J_{N+1\{k_1,k_2+1,k_3\}}+J_{N+1\{k_1,k_2,k_3+1\}} 
\end{align}
and
\begin{align}
\label{k3_up}
&(\Delta_t+k_t-N-2d)J_{N\{k_1,k_2,k_3\}} \nonumber\\
&\qquad=-p_1^2J_{N+1\{k_1-1,k_2,k_3\}}-p_2^2J_{N+1\{k_1,k_2-1,k_3\}}-p_3^2J_{N+1\{k_1,k_2,k_3-1\}}\,.
\end{align}
Finally, we provide the action of the differential operator $\Xi$ given in Eq.~\eqref{Xi_def} on triple-$K$ integrals:
\begin{align}
&(\Xi+a) J_{N\{k_1,k_2,k_3\}}
\nonumber
\\
&=\left(\frac{-\nu_1+\nu_2+\De_3-s}{2}-k_1+k_2+a\right)J_{N\{k_1,k_2,k_3\}}
\nonumber
\\
&\quad
-(\nu_1+k_1+1)J_{N\{k_1+1,k_2,k_3-1\}}+(\nu_2+k_2+1)J_{N\{k_1,k_2+1,k_3-1\}}
\nonumber
\\
&\quad
+\frac{J_{N+1\{k_1+1,k_2,k_3\}}+J_{N+1\{k_1+2,k_2,k_3-1\}}-J_{N+1\{k_1,k_2+1,k_3\}}-J_{N+1\{k_1,k_2+2,k_3-1\}}}{2}
\nonumber
\\
&=\left(\frac{-\De_1+\De_2+\De_3-s}{2}-k_1+k_2+a\right)(\De_t+k_t+N-d-2)J_{N-1\{k_1,k_2,k_3-1\}}
\nonumber
\\
&\quad
+\frac{k_1-k_2+k_3+N+s-2-2a}{2}J_{N\{k_1+1,k_2,k_3-1\}}
\nonumber
\\
&\quad
+\frac{k_1-k_2-k_3-N+s-2\Delta_3+2-2a}{2}J_{N\{k_1,k_2+1,k_3-1\}}
\label{Xi_on_tripleK}
\,,
\end{align}
where we used Eq.~\eqref{k3_down} at the second equality.

\subsection{Zero-momentum limit}
\label{subsec:zero}

When solving the secondary WT identities in the main text, we use the zero-momentum limit, $\bsp_3\rightarrow 0$, of triple-$K$ integrals. In this limit, triple-$K$ integrals reduce to monomials of $p=p_1=p_2$. We write its coefficient as $j_{N\{k_1,k_2,k_3\}}$:
\begin{align}
&j_{N\{k_1,k_2,k_3\}}=p^{-\Delta_t-k_t+2d+N}\lim_{\bsp_3\rightarrow 0}J_{N\{k_1,k_2,k_3\}} \nn\\*
&=\frac{2^{\frac{d}{2}-1+N}\Gamma\left(\Delta_{3}-\frac{d}{2}+k_{3}\right)}{\Gamma\left(d-\Delta_{3}-k_{3}+N\right)}\prod_{u,v=\pm 1}\Gamma\left(\frac{d-\Delta_{3}-k_{3}+N+u\left(\nu_{1}+k_{1}\right)+v\left(\nu_{2}+k_{2}\right)}{2}\right) \,. \label{zero-p3}
\end{align}
To derive this expression, we have used 
\begin{align}
\lim_{x\rightarrow 0}K_\nu (x)=\frac{\Gamma (\nu)2^{\nu -1}}{x^\nu}\quad \text{for}\quad\nu\,\cancel{\in}\,\mathbb{Z} \quad \text{and}\quad\nu > 0 \,,
\end{align}
and the formula~\cite{GR},
\begin{align}
\int_0^{\infty}dx\;x^{-\lambda} K_{\mu}(ax)K_{\nu}(bx)
&=\frac{2^{-2-\lambda}a^{-\nu+\lambda-1}b^\nu}{\Gamma (1-\lambda)}\prod_{u,v=\pm 1}\Gamma\left(\frac{1-\lambda +u\mu+v\nu}{2}\right) \nonumber\\
&\times {}_2F_1\left(\frac{1-\lambda+\mu+\nu}{2},\frac{1-\lambda-\mu+\nu}{2};1-\lambda;1-\frac{b^2}{a^2}\right) \,,
\end{align}
for ${\rm Re}\,(a+b)>0$ and ${\rm Re}\,\lambda<1-|{\rm Re}\,\mu|-|{\rm Re}\,\nu|$, Here ${}_2F_1(a,b;c;x)$ is Gauss's hypergeometric function,
\begin{align}
{}_2F_1(a,b;c;x)=\sum_{n=0}^{\infty}\frac{(a)_n(b)_n}{(c)_n}\frac{x^n}{n!} \,.
\end{align}

\section{Another derivation of the closed form}
\label{app:another}

In this appendix we provide another derivation of the closed form of three-point functions, $\lan \vphi_1(\bsp_1)\vphi_2(\bsp_2)\ep_3^s.O(\bsp_3) \ran'$, $\lan \vphi(\bsp_1)\ep_2.J(\bsp_2)\ep_3^s.O(\bsp_3) \ran'$, and $\lan \vphi(\bsp_1)\ep_2^2.T(\bsp_2)\ep_3^s.O(\bsp_3) \ran'$.

\subsection{Two scalars and one tensor}

Let us begin with three-point functions $\lan \vphi_1(\bsp_1)\vphi_2(\bsp_2)\ep_3^s.O(\bsp_3) \ran'$ of two scalars and one tensor. Our starting point is the triple-$K$ expansion,
\begin{align}
A_n=\sum_{k_1,k_2\geq0}a_{n\{k_1,k_2\}}J_{n+k_1+k_2\{k_1,k_2,n-s\}}\,,
\end{align}
of the coefficient function $A_n$ ($0\leq n \leq s$) given in Eq.~\eqref{ssO-ansatz}. As we mentioned earlier, the WT identities~\eqref{ssO-P1}-\eqref{ssO-P2} can be rephrased in terms of the coefficients $a_{n\{k_1,k_2\}}$ as
\begin{align}
\label{initialWT_OOS}
k_1a_{n\{k_1,k_2\}}=0\,,
\quad
k_2a_{n\{k_1,k_2\}}=a_{n+1\{k_1,k_2-1\}}\,.
\end{align}
In the main text we showed that the closed form~\eqref{closed_OOS} obtained algebraically indeed satisfies Eq.~\eqref{initialWT_OOS}. Instead, here we use these two conditions to determine $a_n$: The former requires that the nonzero coefficients appear only at $k_1=0$, whereas the latter implies
\begin{align}
\label{a_n_OOS}
a_{n\{0,k\}}=\frac{a_{n+k\{0,0\}}}{k!}\,.
\end{align}
Therefore, our task is now reduced to determining $a_{n\{0,0\}}$. We then use the recursion relation~\eqref{00s-recursion}, taking the form~\eqref{WT_closed} with $\alpha=0$. In Appendix~\ref{subsec:formulae_rec} we solve Eq.~\eqref{WT_closed} for general $\alpha$. Applying the general solution \eqref{k1=0_general}, we find
\begin{align}
a_{s-n\{0,0\}}=\cC_A\frac{2^{n}(1-\tfrac{s-\De_1+\De_2+\De_3}{2})_{n}(1+\tfrac{d-s-\De_t}{2})_{n}}{n!(2-\Delta_3-s)_{n}}\,,
\end{align}
where we used $a_{s\{0,0\}}=\cC_A$. Combining with Eq.~\eqref{a_n_OOS}, we conclude that nonzero coefficients $a_n$ are\footnote{
In contrast to the approach presented in the main text, it is not manifest in this derivation if the full correlator satisfies the WT identity~\eqref{00s-recursion}, which has to be checked separately.}
\begin{align}
a_{s-n\{0,k\}}=\cC_A\frac{2^{n-k}(1\!-\!\tfrac{s-\De_1+\De_2+\De_3}{2})_{n-k}(1+\tfrac{d-s-\De_t}{2})_{n-k}}{k!(n-k)!(2-\Delta_3-s)_{n-k}}
\quad
(0\leq k\leq n)\,.
\end{align}

\subsection{Single spin $1$ conserved current}

We next consider three-point functions, $\lan \vphi(\bsp_1)\ep_2.J(\bsp_2)\ep_3^s.O(\bsp_3) \ran'$, with a spin $1$ conserved current. In this case we expand the coefficient functions in Eq.~\eqref{OJS_ansatz} as
\begin{align}
A_n&=\sum_{k_1,k_2\geq0}a_{n\{k_1,k_2\}}J_{n+k_1+k_2+1\{k_1,k_2,n-s\}}\,,
\\
B_n&=\sum_{k_1,k_2\geq0}b_{n\{k_1,k_2\}}J_{n+k_1+k_2\{k_1,k_2,n-s+1\}}\,.
\end{align}
First, the primary WT identities~\eqref{e2p1bp1}-\eqref{e2e3bp2} are rephrased as
\begin{align}
&k_1a_{n\{k_1,k_2\}}=0\,, 
&&k_2a_{n\{k_1,k_2\}}=a_{n+1\{k_1,k_2-1\}}\,, \label{01s-initial-conseq1}
\\
&k_1b_{n\{k_1,k_2\}}=a_{n+1\{k_1-1,k_2\}}\,,
&&
k_2b_{n\{k_1,k_2\}}=b_{n+1\{k_1,k_2-1\}}\,, \label{01s-initial-conseq2}
\end{align}
which imply that nonzero coefficients are
\begin{align}
\label{ab_nonzero}
a_{n\{0,k\}}=\frac{b_{n-1+k\{1,0\}}}{k!}\,,
\quad
b_{n\{0,k\}}=\frac{b_{n+k\{0,0\}}}{k!}\,,
\quad
b_{n\{1,k\}}=\frac{b_{n+k\{1,0\}}}{k!}\,.
\end{align}
Our task is now to determine $b_{n\{0,0\}}$ and $b_{n\{1,0\}}$. For this purpose, we use the recursion relation~\eqref{01s-Bn}, which is of the form~\eqref{WT_closed} with $\alpha=1$. Applying the general solutions~\eqref{k1=0_general}-\eqref{k1=1_general}, we find
\begin{align}
b_{s-1-n\{0,0\}}&=\cC_B\frac{2^{n}(\frac{1}{2}+\tfrac{\De_1-\De_2-\De_3-s}{2})_{n}(\frac{3}{2}+\frac{d-s-\De_t}{2})_{n}}{n!(2-\Delta_3-s)_{n}}\,,
\\
b_{s-1-n\{1,0\}}&
=\cC_A\frac{2^{n}(\frac{3}{2}+\tfrac{\De_1-\De_2-\De_3-s}{2})_{n}(\frac{1}{2}+\frac{d-s-\De_t}{2})_{n}}{n!(2\!-\!\Delta_3\!-\!s)_{n}}
\nonumber
\\
&\quad
-\cC_B\frac{2^{n-1}(\frac{3}{2}+\tfrac{\De_1-\De_2-\De_3-s}{2})_{n-1}(\frac{3}{2}+\frac{d-s-\De_t}{2})_{n-1}}{(n-1)!(2\!-\!\Delta_3\!-\!s)_{n}}
\,,
\end{align}
where we used $b_{s-1\{0,0\}}=\cC_B$ and $b_{s-1\{1,0\}}=\cC_A$. All the other coefficients are obtained by using Eq.~\eqref{ab_nonzero}.

\subsection{Single energy-momentum tensor}

Finally, let us consider three-point functions, $\lan \vphi(\bsp_1)\ep_2^2.T(\bsp_2)\ep_3^s.O(\bsp_3) \ran'$, with an energy-momentum tensor. First, we expand the coefficient functions in Eqs.~\eqref{OTS_ansatz1}-\eqref{OTS_ansatz3} as
\begin{align}
A_n&=\sum_{k_1,k_2\geq0}a_{n\{k_1,k_2\}}J_{n+k_1+k_2+2\{k_1,k_2,n-s\}}\,,
\\*
B_n&=\sum_{k_1,k_2\geq0}b_{n\{k_1,k_2\}}J_{n+k_1+k_2+1\{k_1,k_2,n-s+1\}}\,,
\\*
C_n&=\sum_{k_1,k_2\geq0}c_{n\{k_1,k_2\}}J_{n+k_1+k_2\{k_1,k_2,n-s+2\}}\,.
\end{align}
Then, the primary WT identities~\eqref{02s-initial1}-\eqref{02s-initial6} are rephrased as
\begin{align}
\label{abc_nonzero1}
&k_1a_{n\{k_1,k_2\}}=0\,,
&&k_2a_{n\{k_1,k_2\}}=a_{n+1\{k_1,k_2-1\}}\,,
\\
\label{abc_nonzero2}
&k_1b_{n\{k_1,k_2\}}=2a_{n+1\{k_1-1,k_2\}}\,,
&&k_2b_{n\{k_1,k_2\}}=b_{n+1\{k_1,k_2-1\}}\,,
\\
\label{abc_nonzero3}
&k_1c_{n\{k_1,k_2\}}=b_{n+1\{k_1-1,k_2\}}\,,
&&
k_2c_{n\{k_1,k_2\}}=c_{n+1\{k_1,k_2-1\}}\,,
\end{align}
which imply that nonzero coefficients are
\begin{align}
c_{n\{0,k\}}&=\frac{c_{n+k\{0,0\}}}{k!}\,,
\\
c_{n\{1,k\}}&
=b_{n+1\{0,k\}}
=\frac{c_{n+k\{1,0\}}}{k!}\,,
\\
c_{n\{2,k\}}&
=\frac{1}{2}b_{n+1\{1,k\}}=a_{n+2\{0,k\}}=\frac{c_{n+k\{2,0\}}}{k!}\,.
\end{align}
Our task is now to determine $c_{n\{0,0\}}$, $c_{n\{1,0\}}$, and $c_{n\{2,0\}}$. For this purpose, we use the recursion relation~\eqref{02s-Cn}, taking the form~\eqref{WT_closed} with $\alpha=2$. Applying the general solutions~\eqref{k1=0_general}-\eqref{k1=2_general}, we find
\begin{align}
c_{s-2-n\{0,0\}}
&=\cC_C\frac{2^n(\tfrac{\De_1-\De_2-\De_3-s}{2})_n(2+\frac{d-s-\De_t}{2})_n}{n!(2\!-\!\Delta_3\!-\!s)_n}
\,,
\\
c_{s-2-n\{1,0\}}&
=\cC_B\frac{2^n(1+\tfrac{\De_1-\De_2-\De_3-s}{2})_n(1+\frac{d-s-\De_t}{2})_n}{n!(2\!-\!\Delta_3\!-\!s)_n}
\nonumber
\\
&\quad
-\cC_C\frac{2^{n}(1+\tfrac{\De_1-\De_2-\De_3-s}{2})_{n-1}(2+\frac{d-s-\De_t}{2})_{n-1}}{(n-1)!(2\!-\!\Delta_3\!-\!s)_n}\,,
\\
c_{s-2-n\{2,0\}}&
=\cC_A\frac{2^n(2+\tfrac{\De_1-\De_2-\De_3-s}{2})_n(\frac{d-s-\De_t}{2})_n}{n!(2\!-\!\Delta_3\!-\!s)_n}
\nonumber
\\
&\quad
-\cC_B
\frac{2^{n-1}(2+\tfrac{\De_1-\De_2-\De_3-s}{2})_{n-1}(1+\frac{d-s-\De_t}{2})_{n-1}}{(n-1)!(2\!-\!\Delta_3\!-\!s)_n}
\nonumber
\\
&\quad
+\cC_C \frac{2^{n-2}(2+\tfrac{\De_1-\De_2-\De_3-s}{2})_{n-2}(2+\frac{d-s-\De_t}{2})_{n-2}}{(n-2)!(2\!-\!\Delta_3\!-\!s)_n}
\,,
\end{align}
where we used $c_{s-2\{0,0\}}=\cC_C$, $c_{s-2\{1,0\}}=\cC_B$, and $c_{s-2\{2,0\}}=\cC_A$. All the other coefficients are obtained by using Eqs.~\eqref{abc_nonzero1}-\eqref{abc_nonzero3}.

\subsection{Useful formulae}
\label{subsec:formulae_rec}

In this section we encountered WT identities of the form,
\begin{align}
\label{WT_closed}
\left(\Xi+n+\frac{3}{2}\alpha\right)F_{n+1}=(-s+n+\alpha)(\Delta_3-1+n+\alpha)F_n\,,
\end{align}
with $F_n$ being a sum over triple-$K$ functions given by
\begin{align}
F_n=\sum_{k_1,k_2\geq0}f_{n\{k_1,k_2\}}J_{n+k_1+k_2\{k_1,k_2,n-s+\alpha\}}\,.
\end{align}
In particular, we are interested in its $k_2=0$ sector. Using the formula~\eqref{Xi_on_tripleK}, we find that Eq.~\eqref{WT_closed} implies
\begin{align}
&f_{s-\alpha-n\{k_1,0\}}
\nonumber
\\
&=2\frac{(1\!-\!\frac{\alpha}{2}+k_1+\tfrac{\De_1-\De_2-\De_3-s}{2}+n\!-\!1)(1+\frac{\alpha}{2}\!-\!k_1+\frac{d-s-\De_t}{2}+n\!-\!1)}{(1+n\!-\!1)(2\!-\!\Delta_3\!-\!s+n\!-\!1)}f_{s-\alpha-n+1\{k_1,0\}}
\nonumber
\\
&\quad
+\frac{(k_1-1-\alpha)}{(1+n\!-\!1)(2\!-\!\Delta_3\!-\!s+n\!-\!1)}f_{s-\alpha-n+1\{k_1-1,0\}}\,.
\end{align}

It is convenient to note that its solution is generally given by
\begin{align}
&f_{s-\alpha-n\{k_1,0\}}
\nonumber
\\
&=\frac{2^n(1\!-\!\frac{\alpha}{2}+k_1+\tfrac{\De_1-\De_2-\De_3-s}{2})_n(1+\frac{\alpha}{2}\!-\!k_1+\frac{d-s-\De_t}{2})_n}{n!(2\!-\!\Delta_3\!-\!s)_n}f_{s-\alpha\{k_1,0\}}
\nonumber
\\
&\quad
+(k_1\!-\!1\!-\!\alpha)\sum_{m=0}^{n-1}\frac{(2\!-\!\frac{\alpha}{2}\!+\!k_1\!+\!\tfrac{\De_1-\De_2-\De_3-s}{2}\!+\!m)_{n-m-1}(2\!+\!\frac{\alpha}{2}\!-\!k_1\!+\!\frac{d-s-\De_t}{2}\!+\!m)_{n-m-1}}{(1+m)_{n-m}(2\!-\!\Delta_3\!-\!s+m)_{n-m}}
\nonumber
\\
\label{general_recursive}
&\qquad\qquad\qquad\qquad\quad
\times 2^{n-m-1}f_{s-\alpha-m\{k_1-1,0\}}\,,
\end{align}
where $f_{s-\alpha-m\{k,0\}}=0$ for $k<0$ in the last line. We can then derive concrete expressions for $f_{s-\alpha-n\{k_1,0\}}$ recursively in $k_1$. First, for $k_1=0$ we have
\begin{align}
\label{k1=0_general}
f_{s-\alpha-n\{0,0\}}
&=\frac{2^n(1\!-\!\frac{\alpha}{2}+\tfrac{\De_1-\De_2-\De_3-s}{2})_n(1+\frac{\alpha}{2}+\frac{d-s-\De_t}{2})_n}{n!(2\!-\!\Delta_3\!-\!s)_n}f_{s-\alpha\{0,0\}}
\,.
\end{align}
Combining this with Eq.~\eqref{general_recursive}, we find
\begin{align}
f_{s-\alpha-n\{1,0\}}
&=\frac{2^n(2\!-\!\frac{\alpha}{2}+\tfrac{\De_1-\De_2-\De_3-s}{2})_n(\frac{\alpha}{2}+\frac{d-s-\De_t}{2})_n}{n!(2\!-\!\Delta_3\!-\!s)_n}f_{s-\alpha\{1,0\}}
\nonumber
\\*
\label{k1=1_general}
&\quad
-\alpha\frac{2^{n-1}(2\!-\!\frac{\alpha}{2}+\tfrac{\De_1-\De_2-\De_3-s}{2})_{n-1}(1+\frac{\alpha}{2}+\frac{d-s-\De_t}{2})_{n-1}}{(n-1)!(2\!-\!\Delta_3\!-\!s)_n}f_{s-\alpha\{0,0\}}
\,,
\end{align}
where we used $(a+m)_{n-m}(a)_m=(a)_n$ and $\displaystyle\sum_{m=0}^{n-1}\frac{1}{(a+m)(a+m+1)}=\frac{n}{a(a+n)}$. Similarly, we arrive at
\begin{align}
f_{s-\alpha-n\{2,0\}}
&=\frac{2^n(3\!-\!\frac{\alpha}{2}+\tfrac{\De_1-\De_2-\De_3-s}{2})_n(-1+\frac{\alpha}{2}+\frac{d-s-\De_t}{2})_n}{n!(2\!-\!\Delta_3\!-\!s)_n}f_{s-\alpha\{2,0\}}
\nonumber
\\*
&\quad
-(\alpha\!-\!1)
\frac{2^{n-1}(3\!-\!\frac{\alpha}{2}+\tfrac{\De_1-\De_2-\De_3-s}{2})_{n-1}(\frac{\alpha}{2}+\frac{d-s-\De_t}{2})_{n-1}}{(n-1)!(2\!-\!\Delta_3\!-\!s)_n}f_{s-\alpha\{1,0\}}
\nonumber
\\*
\label{k1=2_general}
&\quad
+\alpha(\alpha\!-\!1)\frac{2^{n-3}(3\!-\!\frac{\alpha}{2}+\tfrac{\De_1-\De_2-\De_3-s}{2})_{n-2}(1+\frac{\alpha}{2}+\frac{d-s-\De_t}{2})_{n-2}}{(n-2)!(2\!-\!\Delta_3\!-\!s)_n}f_{s-\alpha\{0,0\}}
\,,
\end{align}
where we used $\displaystyle\sum_{m=0}^{n-1}\frac{m}{(a+m-1)(a+m)(a+m+1)}=\frac{n(n-1)}{2a(a+n)(a+n-1)}$.

\section{Correlators with two energy-momentum tensors}
\label{app:TTs}
This appendix summarizes the results for correlators with two energy-momentum tensors:
\begin{align}
\lan \ep_1^{2}.T(\bsp_1) \ep_2^{2}.T(\bsp_2) \ep_3^s.O(\bsp_3) \ran' \,.
\end{align}
The strategy is parallel to the two spin $1$ conserved current case discussed in Sec.~\ref{sec:TTs}. First, we perform tensor decomposition to write the correlator as a sum of $14$ terms as
\begin{alignat}{2}
&\lan \ep_1^2.T(\bsp_1) \ep_2^2.T(\bsp_2) \ep_3^s.O(\bsp_3) \ran' \nn\\
&= 
(\bsep_1^2\cdot\boldsymbol{\Pi}_1\cdot \bsp_2^2)(\bsep_2^2\cdot\boldsymbol{\Pi}_2\cdot \bsp_1^2)X_A 
&&+(\bsep_1^2\cdot\boldsymbol{\Pi}_1\cdot \bsp_2\bsep_3)(\bsep_2^2\cdot\boldsymbol{\Pi}_2\cdot \bsp_1\bsep_3)X_B \nn\\
&\quad
+(\bsep_1^2\cdot\boldsymbol{\Pi}_1\cdot \bsep_3^2)(\bsep_2^2\cdot\boldsymbol{\Pi}_2\cdot \bsep_3^2) X_C 
&&+(\bsep_1^2\cdot\boldsymbol{\Pi}_1\cdot \bsp_2)_\mu(\bsep_2^2\cdot\boldsymbol{\Pi}_2\cdot \bsp_1)^\mu X_D \nn\\
&\quad
+(\bsep_1^2\cdot\boldsymbol{\Pi}_1\cdot \bsep_3)_\mu(\bsep_2^2\cdot\boldsymbol{\Pi}_2\cdot \bsep_3)^\mu X_E
&&+(\bsep_1^2\cdot\boldsymbol{\Pi}_1)_{\mu\nu}(\bsep_2^2\cdot\boldsymbol{\Pi}_2)^{\mu\nu}X_F \nn\\
&\quad
+(\bsep_1^2\cdot\boldsymbol{\Pi}_1\cdot \bsp_2^2)(\bsep_2^2\cdot\boldsymbol{\Pi}_2\cdot \bsp_1\bsep_3) X_G
&&+(\bsep_1^2\cdot\boldsymbol{\Pi}_1\cdot \bsp_2\bsep_3)(\bsep_2^2\cdot\boldsymbol{\Pi}_2\cdot \bsp_1^2) Y_G \nn\\
&\quad
+(\bsep_1^2\cdot\boldsymbol{\Pi}_1\cdot \bsp_2^2)(\bsep_2^2\cdot\boldsymbol{\Pi}_2\cdot \bsep_3^2)X_H
&&+(\bsep_1^2\cdot\boldsymbol{\Pi}_1\cdot \bsep_3^2)(\bsep_2^2\cdot\boldsymbol{\Pi}_2\cdot \bsp_1^2)Y_H \nn\\
&\quad
+(\bsep_1^2\cdot\boldsymbol{\Pi}_1\cdot \bsp_2\bsep_3)(\bsep_2^2\cdot\boldsymbol{\Pi}_2\cdot \bsep_3^2)X_I
&&+(\bsep_1^2\cdot\boldsymbol{\Pi}_1\cdot \bsep_3^2)(\bsep_2^2\cdot\boldsymbol{\Pi}_2\cdot \bsp_1\bsep_3)Y_I \nn\\
&\quad
+(\bsep_1^2\cdot\boldsymbol{\Pi}_1\cdot \bsp_2)_\mu(\bsep_2^2\cdot\boldsymbol{\Pi}_2\cdot \bsep_3)^\mu X_J
&\;&+(\bsep_1^2\cdot\boldsymbol{\Pi}_1\cdot \bsep_3)_\mu(\bsep_2^2\cdot\boldsymbol{\Pi}_2\cdot \bsp_1)^\mu Y_J\,.
\label{22s-ansatz}
\end{alignat}
Note that $X_C$, $X_I$ and $Y_I$ do not exist when $s=2$ because we have only two $\bsep_3$. We then expand each term in $\xi=\bsep\cdot\bsp_2$ and $\zt=\bsep\cdot(\bsp_1+\bsp_2)$ as
\begin{align}
&X_A = \sum_{n=0}^{s}\frac{1}{n!}\xi^{n}\zeta^{s-n}A_n \,,
&&X_B = \sum_{n=0}^{s-2}\frac{1}{n!}\xi^{n}\zeta^{s-n}B_n \,, \nn\\
&X_C =\sum_{n=0}^{s-4}\frac{1}{n!}\xi^{n}\zeta^{s-n-4}C_n \,,
&&X_D =\sum_{n=0}^{s}\frac{1}{n!}\xi^{n}\zeta^{s-n}D_n \,, \nn\\
&X_E =\sum_{n=0}^{s-2}\frac{1}{n!}\xi^{n}\zeta^{s-n-2}E_n \,,
&&X_F =\sum_{n=0}^{s}\frac{1}{n!}\xi^{n}\zeta^{s-n}F_n \,, \nn\\
&X_G =\sum_{n=0}^{s-1}\frac{1}{n!}\xi^{n}\zeta^{s-n-1}G_n \,,
&&Y_G =\sum_{n=0}^{s-1}\frac{1}{n!}\xi^{n}\zeta^{s-n-1}G_n^\star \,, \nn\\
&X_H =\sum_{n=0}^{s-2}\frac{1}{n!}\xi^{n}\zeta^{s-n-2}H_n \,,
&&Y_H =\sum_{n=0}^{s-2}\frac{1}{n!}\xi^{n}\zeta^{s-n-2}H_n^\star \,,\nn\\
&X_I =\sum_{n=0}^{s-3}\frac{1}{n!}\xi^{n}\zeta^{s-n-3}I_n \,,
&&Y_I =\sum_{n=0}^{s-3}\frac{1}{n!}\xi^{n}\zeta^{s-n-3}I_n^\star \,,\nn\\
&X_J =\sum_{n=0}^{s-1}\frac{1}{n!}\xi^{n}\zeta^{s-n-1}J_n \,,
&&Y_J =\sum_{n=0}^{s-1}\frac{1}{n!}\xi^{n}\zeta^{s-n-1}J_n^\star \,.  
\end{align}
Note that the exchange symmetry $\ep_1,p_1\leftrightarrow\ep_2,p_2$ implies for example
\begin{align} 
&A_s(p_1,p_2,p_3)=(-1)^sA_s(p_2,p_1,p_3)\,,
&&B_{s-2}(p_1,p_2,p_3)=(-1)^{s-2}B_{s-2}(p_2,p_1,p_3) \nonumber\,,\\
&C_{s-4}(p_1,p_2,p_3)=(-1)^{s-4}C_{s-4}(p_2,p_1,p_3)\,,
&&D_{s}(p_1,p_2,p_3)=(-1)^{s}D_{s}(p_2,p_1,p_3) \nonumber\,,\\
&E_{s-2}(p_1,p_2,p_3)=(-1)^{s-2}E_{s-2}(p_2,p_1,p_3)\,,
&&F_{s}(p_1,p_2,p_3)=(-1)^{s}F_{s}(p_2,p_1,p_3) \,,\\*
&G_{s-1}(p_1,p_2,p_3)=(-1)^{s-1}G^\star_{s-1}(p_2,p_1,p_3)\,,
&&H_{s-2}(p_1,p_2,p_3)=(-1)^{s-2}H^\star_{s-2}(p_2,p_1,p_3) \nonumber\,,\\
&I_{s-3}(p_1,p_2,p_3)=(-1)^{s-3}I^\star_{s-3}(p_2,p_1,p_3)\,,
&&J_{s-1}(p_1,p_2,p_3)=(-1)^{s-1}J^\star_{s-1}(p_2,p_1,p_3) \nonumber\,.
\end{align}
Based on this ansatz, we solve the special conformal WT identity,
\begin{align}
0 = (\bsb\cdot\bsK_s +\bsb\cdot\bsK_{\ep_1}+ \bsb\cdot\bsK_{\ep_2})
\lan \ep_1^{2}.T(\bsp_2) \ep_2^{2}.T(\bsp_2) \ep_3^s.O(\bsp_3) \ran' \,,
\end{align}
as well as the dilatation WT identity, which yields the homogeneity conditions. 
\begin{align}
A_n(\lm p_1,\lm p_2,\lm p_3) &= \lm^{\De_t-2d-s-4} A_n(p_1,p_2,p_3) \,, \nn\\
B_n(\lm p_1,\lm p_2,\lm p_3) &= \lm^{\De_t-2d-s} B_n(p_1,p_2,p_3) \,, \nn\\
C_n(\lm p_1,\lm p_2,\lm p_3) &= \lm^{\De_t-2d-s+4} C_n(p_1,p_2,p_3) \,, \nn\\
D_n(\lm p_1,\lm p_2,\lm p_3) &= \lm^{\De_t-2d-s-2} D_n(p_1,p_2,p_3) \,, \nn\\
E_n(\lm p_1,\lm p_2,\lm p_3) &= \lm^{\De_t-2d-s+2} E_n(p_1,p_2,p_3) \,, \nn\\
F_n(\lm p_1,\lm p_2,\lm p_3) &= \lm^{\De_t-2d-s} F_n(p_1,p_2,p_3) \,, \nn\\
G_n(\lm p_1,\lm p_2,\lm p_3) &= \lm^{\De_t-2d-s-2} G_n(p_1,p_2,p_3) \,, \label{TTS-hom} \\
G^*_n(\lm p_1,\lm p_2,\lm p_3) &= \lm^{\De_t-2d-s-2} G^*_n(p_1,p_2,p_3) \,, \nn\\
H_n(\lm p_1,\lm p_2,\lm p_3) &= \lm^{\De_t-2d-s} H_n(p_1,p_2,p_3) \,, \nn\\
H^*_n(\lm p_1,\lm p_2,\lm p_3) &= \lm^{\De_t-2d-s} H^*_n(p_1,p_2,p_3) \,, \nn\\
I_n(\lm p_1,\lm p_2,\lm p_3) &= \lm^{\De_t-2d-s+2} I_n(p_1,p_2,p_3) \,, \nn\\
I^*_n(\lm p_1,\lm p_2,\lm p_3) &= \lm^{\De_t-2d-s+2} I^*_n(p_1,p_2,p_3) \,, \nn\\
J_n(\lm p_1,\lm p_2,\lm p_3) &= \lm^{\De_t-2d-s} J_n(p_1,p_2,p_3) \,, \nn\\
J^*_n(\lm p_1,\lm p_2,\lm p_3) &= \lm^{\De_t-2d-s} J^*_n(p_1,p_2,p_3) \,. \nn
\end{align}

\subsection{List of primary and secondary WT identities}

The special conformal WT identities are decomposed into $42$ primary WT identities and $16$ secondary WT identities. First, $28$ of the primary ones are associated to $\bsb\cdot\bsp_1$ and $\bsb\cdot\bsp_2$:

\begin{align}
&0=\left(P_1^{(-4,0)}+8p_3^{-2}\te_3\right)X_A \,, ,
&&0=\left(P_2^{(-4,0)}+8p_3^{-2}\te_3\right)X_A \,, \nonumber\\
&0=\left(P_1^{(0,2)}+4p_3^{-2}\te_3\right)X_B +4\pd_\xi Y_G\,,
&&0=\left(P_2^{(0,2)}+4p_3^{-2}\te_3\right)X_B -4\pd_\xi X_G\,, \nonumber\\
&0=P_1^{(4,4)}X_C+2\pd_\xi Y_I
&&0=P_2^{(4,4)}X_C-2\pd_\xi X_I \,, \nonumber\\
&0=\left(P_1^{(-2,0)}+4p_3^{-2}\te_3\right)X_D +8X_A\,, 
&&0=\left(P_2^{(-2,0)}+4p_3^{-2}\te_3\right)X_D +8X_A\,, \nonumber\\
&0=P_1^{(2,2)}X_E+2X_B +2\pd_\xi Y_J\,,
&&0=P_2^{(2,2)}X_E+2X_B -2\pd_\xi X_J\,,\nonumber\\
&0=P_1^{(0,0)}X_F+2X_D \,,
&&0=P_2^{(0,0)}X_F+2X_D \,, \nonumber\\
&0=\left(P_1^{(-2,1)}+6p_3^{-2}\te_3\right)X_G+4\pd_\xi X_A\,,
&&0=\left(P_2^{(-2,1)}+6p_3^{-2}\te_3\right)X_G\,,
\label{TTS_initial}\\
&0=\left(P_1^{(-2,1)}+6p_3^{-2}\te_3\right)Y_G\,,
&&0=\left(P_2^{(-2,1)}+6p_3^{-2}\te_3\right)Y_G-4\pd_\xi X_A\,,\nonumber\\
&0=\left(P_1^{(0,2)}+4p_3^{-2}\te_3\right)X_H+2\pd_\xi X_G\,,
&&0=\left(P_2^{(0,2)}+4p_3^{-2}\te_3\right)X_H\,,
\nonumber\\
&0=\left(P_1^{(0,2)}+4p_3^{-2}\te_3\right)Y_H\,,
&&0=\left(P_2^{(0,2)}+4p_3^{-2}\te_3\right)Y_H-2\pd_\xi Y_G\,,\nonumber\\
&0=\left(P_1^{(2,3)}+2p_3^{-2}\te_3\right)X_I+2\pd_\xi X_B\,,
&&0=\left(P_2^{(2,3)}+2p_3^{-2}\te_3\right)X_I-4\pd_\xi X_H\,,\nonumber\\
&0=\left(P_1^{(2,3)}+2p_3^{-2}\te_3\right)Y_I+4\pd_\xi Y_H\,,
&&0=\left(P_2^{(2,3)}+2p_3^{-2}\te_3\right)Y_I-2\pd_\xi X_B\,,\nonumber\\
&0=\left(P_1^{(0,1)}+2p_3^{-2}\te_3\right)X_J+2\pd_\xi X_D+4X_G\,,
&&0=\left(P_2^{(0,1)}+2p_3^{-2}\te_3\right)X_J+4X_G\,,
\nonumber\\
&0=\left(P_1^{(0,1)}+2p_3^{-2}\te_3\right)Y_J+4Y_G\,,
&&0=\left(P_2^{(0,1)}+2p_3^{-2}\te_3\right)Y_J-2\pd_\xi X_D+4Y_G\,.
\nonumber
\end{align}
The other $14$ are associated with $\bsb\cdot\bsep_3$:
\begin{align}
0&=\left(R^{(-4,0)}+4\pd_\xi+8\pd_\zt\right)X_A+2p_3^{-2}\te_3(X_G+Y_G) \,, \nonumber\\
0&=\left(R^{(0,2)}+4\pd_\xi+8\pd_\zeta\right)X_B+4p_3^{-2}\te_3(X_I+Y_I) \,, \nonumber\\
0&=\left(R^{(4,4)}+4\pd_\xi+8\pd_\zeta\right)X_C\,, \nonumber\\
0&=\left(R^{(-2,0)}+2\pd_\xi+4\pd_\zeta\right)X_D+4(X_G+Y_G)+2p_3^{-2}\te_3(X_J+Y_J) \,,\nonumber\\
0&=\left(R^{(2,0)}+2\pd_\xi+4\pd_\zt\right)X_E+4(X_I+Y_I) \,, \nonumber\\
0&=R^{(0,0)}X_F+2(X_J+Y_J) \,, \nonumber\\
0&=\left(R^{(-2,1)}+6\pd_\xi+8\pd_\zeta \right)X_G +2p^{-2}_3\te_3 X_B+4p^{-2}_3\te_3 X_H\,,
\label{TTS_recursion} \\
0&=\left(R^{(-2,1)}+2\pd_\xi+8\pd_\zeta \right)Y_G  +2p^{-2}_3\te_3 X_B+4p^{-2}_3\te_3 Y_H\,,\nonumber\\
0&=\left(R^{(0,2)}+8\pd_\xi+8\pd_\zeta\right)X_H+2p_3^{-2}\te_3 X_I\,,\nonumber\\
0&=\left(R^{(0,2)}+8\pd_\zeta\right)Y_H+2p_3^{-2}\te_3 Y_I\,,\nonumber\\
0&=\left(R^{(2,3)}+6\pd_\xi + 8\pd_\zeta\right)X_I+4p_3^{-2}\te_3 X_C\,,\nonumber\\
0&=\left(R^{(2,3)}+2\pd_\xi + 8\pd_\zeta\right)X_I+4p_3^{-2}\te_3 X_C\,,\nonumber\\
0&=\left(R^{(0,1)}+4\pd_\xi+4\pd_\zeta\right)X_J+2X_B +2p_3^{-2}\te_3 X_E+8 X_H\,,\nonumber\\
0&=\left(R^{(0,1)}+4\pd_\zeta\right)Y_J+2X_B +2p_3^{-2}\te_3 X_E+8 Y_H\,. \nn
\end{align}
To write down the secondary WT identities, it is convenient to introduce
\begin{align}
\hL &=-\frac{\bsp_1\cdot\bsp_2}{p_1^2}\te_1+\te_2+\te_\xi-\Delta_2 +1+d+d\frac{\bsp_1\cdot\bsp_2}{p_1^2}\,,\\*
\hS &=\frac{\xi-\zeta}{p_1^2}\left(d-\te_1\right) +\frac{\zeta}{p_3^2}\te_3\,,\\*
\hs &=\frac{1}{p_1^2}(d-\te_1)\,,
\end{align}
which become simpler in the zero momentum limit $\bsp_3\rightarrow 0$, that is $p_1=p_2=p$ and $\zeta=0$,
\begin{align}
\hL &= \theta_1 \,_{|p_1\rightarrow p}+\theta_2 \,_{|p_2\rightarrow p}+\xi\pd_\xi-d+1 \,,\nonumber\\*
\hS &= \xi\;\hs = \frac{\xi}{p^2}(d-\theta_1 \,_{|p_1\rightarrow p}) \,,
\end{align}
where we set $\Delta_2=d$.
Half of the $16$ secondary WT identities are associated with $(\bsb\cdot\boldsymbol{\Pi}_1\cdot\bsep_1)_{\mu\nu}$ and of the form,
\begin{align}
0&=2\hL X_A +\hs X_D-\hS Y_G\,, \label{22s-second1}\\
0&=-\hS X_B+\pd_\xi X_D +2\hL X_G -2Y_G +\hs X_J \,, \label{22s-second2}\\
0&=-X_B+2\hL X_H-\hS X_I+\pd_\xi X_J\,, \label{22s-second3}\\
0&=\hL X_B+\hs X_E -4Y_H-2\hS Y_I +\pd_\xi Y_J\,, \label{22s-second4}\\
0&=-2\hS X_C +\pd_\xi X_E+\hL X_I -2Y_I\,, \label{22s-second5}\\
0&=\hL Y_G-2\hS Y_H+\hs Y_J\,,\label{22s-second6}\\
0&=\hL X_D+2\hs X_F-\hS Y_J\,, \label{22s-second7}\\
0&=-\hS X_E+2\pd_\xi X_F+\hL X_J-Y_J \,.\label{22s-second8}
\end{align}
Note that the condition~\eqref{22s-second6} does not exits for $s=2$ because it requires $3$ or more $\bsep_3$. The other $8$ equations are associated with $(\bsb\cdot\boldsymbol{\Pi}_2\cdot\bsep_2)_{\mu\nu}$, but they are equivalent to 
the above 8 identities
because of the $1\leftrightarrow2$ exchange symmetry.

\subsection{Solutions for even $s$}

Similarly to the spin $1$ current case in Sec.~\ref{sec:TTs}, it is easy to show that three-point functions vanish when the tensor $O$ has an odd spin $s$. We therefore focus on the even spin case.

\subsubsection*{Initial conditions}

We solve the $\mathcal{O}(\zeta^0)$ terms of Eq.~\eqref{TTS_initial} to find the initial conditions:
\begin{align}
A_s&=\cC_A J_{s+4\{0,0,0\}}\,,\nonumber\\
B_{s-2}&=-4\cC_A J_{s+2\{1,1,0\}}+2\cC_G J_{s+1\{0,0,1\}}+\cC_B J_{s\{0,0,0\}}\,,\nonumber\\
C_{s-4}&=\cC_A J_{s\{2,2,0\}}  -\cC_G \left(2J_{s-1\{1,1,1\}}+J_{s-1\{1,2,0\}}+J_{s-1\{2,1,0\}}\right)\nn\\
&\quad -\left(\cC_B J_{s-2\{1,1,0\}}-\cC_H J_{s-2\{2,0,0\}}-\cC_H J_{s-2\{0,2,0\}}\right)  +\cC_I J_{s-3\{0,0,1\}} +\cC_C J_{s-4\{0,0,0\}}\,,\nonumber\\
D_s&=-4\cC_A J_{s+3\{0,0,1\}}+\cC_D J_{s+2\{0,0,0\}}\,,\nonumber\\
E_{s-2}&=4\cC_A J_{s+1\{1,1,1\}}-\cC_D J_{s\{1,1,0\}}-2\cC_G J_{s\{0,0,2\}} +(\cC_J-\cC_B ) J_{s-1\{0,0,1\}}+\cC_E J_{s-2\{0,0,0\}}\,,
\nonumber\\
F_s&=2\cC_A J_{s+2\{0,0,2\}}-\cC_D J_{s+1\{0,0,1\}}+\cC_F J_{s\{0,0,0\}}\,,\nonumber\\
G_{s-1}&=-G^{\star}_{s-1}=2\cC_A J_{s+3\{1,0,0\}}+\cC_G J_{s+2\{0,0,0\}}\,,\nn\\
H_{s-2}&=H^{\star}_{s-2}
=\cC_A J_{s+2\{2,0,0\}} +\cC_G J_{s+1\{1,0,0\}}+\cC_H J_{s\{0,0,0\}}\,,\nonumber\\
I_{s-3}&=-I^{\star}_{s-3}
=-2\cC_A J_{s+1\{2,1,0\}}+\cC_G\left(2J_{s\{1,0,1\}}+J_{s\{2,0,0\}}\right) \nonumber\\
&\quad\quad\quad\quad\quad
+\cC_B J_{s-1\{1,0,0\}}-2\cC_H J_{s-1\{0,1,0\}}+\cC_I J_{s-2\{0,0,0\}}\,,\nonumber\\
J_{s-1}&=-J^{\star}_{s-1}
=-4\cC_A J_{s+2\{1,0,1\}}+\cC_D J_{s+1\{1,0,0\}}-2\cC_G J_{s+1\{0,0,1\}}+\cC_J J_{s\{0,0,0\}}\,.
\end{align}

\subsubsection*{Recursion relations}

On the other hand, recursion relations follow from Eq.~\eqref{TTS_recursion}. For notational simplicity, we introduce differential operators $\bbD_{(a,b,c)}$ and $\wh\bbD{}_{t(a,b,c)}$ as
\begin{align}
\bbD_{(a,b,c)} &= \frac{(\Xi+n+a)_{s-n-b}}{(-s+n+b)_{s-n-b}(\De_3-1+n+c)_{s-n-b}},\\
\wh\bbD{}_{t(a,b,c)} &= \frac{(\Xi+n+a)_t}{(-s+n+b)_{t+1}(\De_3-1+n+c)_{t+1}}\,.
\end{align}
Using this notation, the solutions for recursion relations are given by
\begin{align}
A_n &=\bbD_{(0,0,0)}+\sum_{t=0}^{s-1-n}\wh\bbD{}_{t(0,0,0)}p_3^{-2}\te_3(G_{n+t}+G^{\star}_{n+t})\,,\nonumber\\
B_n &=\bbD_{(2,2,2)}B_{s-2}+\sum_{t=0}^{s-3-n}\wh\bbD{}_{t(2,2,2)}p_3^{-2}\te_3(I_{n+t}+I^{\star}_{n+t})\,,\nonumber\\
C_n &=\bbD_{(2,4,4)}C_{s-4}\,,\nonumber\\
D_n &=\bbD_{(0,0,0)}D_{s} + \sum_{t=0}^{s-1-n}\wh\bbD{}_{t(0,0,0)}(2G_{n+t}+2G^{\star}_{n+t}+p_3^{-2}\te_3J_{n+t}+p_3^{-2}\te_3J^{\star}_{n+t})\,,\nonumber\\
E_n &=\bbD_{(2,2,2)}E_{s-2} + 2\sum_{t=0}^{s-3-n}\wh\bbD{}_{t(2,2,2)}(I_{n+t}+I^{\star}_{n+t})\,,\nonumber\\
F_n &=\bbD_{(0,0,0)}F_{s} + \sum_{t=0}^{s-1-n}\wh\bbD{}_{t(0,0,0)}p_3^{-2}\te_3(J_{n+t}+J^{\star}_{n+t})\,,\nonumber\\
G_n &=\bbD_{(2,1,1)}G_{s-1}+ \sum_{t=0}^{s-2-n}\wh\bbD{}_{t(2,1,1)}p_3^{-2}\te_3(B_{n+t}+2H_{n+t})\,,\nonumber\\
G^{\star}_n &=\bbD_{(0,1,1)}G^\star_{s-1}+ \sum_{t=0}^{s-2-n}\wh\bbD{}_{t(0,1,1)}p_3^{-2}\te_3(B_{n+t}+2H^{\star}_{n+t})\,,\nonumber\\
H_n &=\bbD_{(4,2,2)}H_{s-2}+\sum_{t=0}^{s-3-n}\wh\bbD{}_{t(4,2,2)}p_3^{-2}\te_3 I_{n+t}\,,\nonumber\\
H^{\star}_n &=\bbD_{(0,2,2)}H^{\star}_{s-2}+\sum_{t=0}^{s-3-n}\wh\bbD{}_{t(0,2,2)}p_3^{-2}\te_3 I^{\star}_{n+t}\,,\nonumber\\
I_n &=\bbD_{(4,3,3)}I_{s-2}+2\sum_{t=0}^{s-4-n}\wh\bbD{}_{t(4,3,3)}p_3^{-2}\te_3 C_{n+t}\,,\nonumber\\
I^{\star}_n &=\bbD_{(2,3,3)}I^{\star}_{s-2}+2\sum_{t=0}^{s-4-n}\wh\bbD{}_{t(2,3,3)}p_3^{-2}\te_3 C_{n+t}\,,\nonumber\\
J_n &=\bbD_{(2,1,1)}J_{s-1} + \sum_{t=0}^{s-2-n}\wh\bbD{}_{t(2,1,1)}(B_{n+t}+p_3^{-2}\te_3E_{n+t}+4H_{n+t})\,,\nonumber\\
J^{\star}_n &=\bbD_{(0,1,1)}B_{s-1} + \sum_{t=0}^{s-2-n}\wh\bbD{}_{t(0,1,1)}(B_{n+t}+p_3^{-2}\te_3E_{n+t}+4H^{\star}_{n+t})\,.
\end{align}

\subsubsection*{Secondary WT identities}

So far we have $10$ free parameters $\cC_A,\ldots,\cC_J$. While there are $8$ secondary WT identities, it turns out that there is one degeneracy among them when applied to the solutions for the primary WT identities. As a result, we are left with $3$ free parameters for $s\geq4$. As we mentioned, the operators $X_C$, $X_I$ and $Y_I$, and the secondary WT identity~\eqref{22s-second6} do not exist for $s=2$, while there still exists one degeneracy. Correspondingly, there remain $2$ free parameters for $s=2$.

\bibliography{WT}{}
\bibliographystyle{utphys}

\end{document}